\documentclass[runningheads]{llncs}

\newcommand{\techrep}{}

\usepackage[paperwidth=155mm,
            paperheight=235mm,
            left=16mm,
            right=16mm,
            top=17mm,
            bottom=25mm
           ]{geometry}

\usepackage{sosy-paper}
\usepackage[textsize=tiny]{todonotes}
\usepackage{wrapfig}

\definetool{\moxi}{MoXI}
\definetool{\moxichecker}{MoXIchecker}
\definetool{\moximcflow}{MoXI-MC-Flow}
\definetool{\pysmt}{PySMT}
\definetool{\btormc}{BtorMC}
\definetool{\msat}{MathSAT}
\definetool{\pono}{Pono}
\definetool{\verilog}{Verilog}
\definetool{\boogie}{Boogie}

\newcommand{\init}{I}
\newcommand{\trans}{T}
\newcommand{\inv}{Inv}
\newcommand{\reachable}{Q}
\newcommand\plotpath{eval-results/tex}
\def\CC{{C\nolinebreak[4]\hspace{-.05em}\raisebox{.4ex}{\tiny\bf ++}}\xspace}

\newcommand{\qfbv}{\texttt{QF\_BV}\xspace}
\newcommand{\qfabv}{\texttt{QF\_ABV}\xspace}
\newcommand{\qflia}{\texttt{QF\_LIA}\xspace}
\newcommand{\qflra}{\texttt{QF\_LRA}\xspace}
\newcommand{\qfnia}{\texttt{QF\_NIA}\xspace}
\newcommand{\qfnra}{\texttt{QF\_NRA}\xspace}
\newcommand{\moxicheckerurl}{https://gitlab.com/sosy-lab/software/moxichecker}
\newcommand{\moxiextask}[2]{\href{\moxicheckerurl/-/blob/main/examples/#1.moxi.json}{#2}}
\definetool{\benchexectoolurl}{\href{https://github.com/sosy-lab/benchexec}{BenchExec}}

\lstdefinelanguage{moxi}{
    alsoletter={-, :},
    keywords=[1]{set-logic, define-system, check-system},
    keywordstyle=[1]\color{violet},
    keywords=[2]{:input, :output, :local, :init, :trans, :inv, :reachable, :query},
    keywordstyle=[2]\color{blue},
    keywords=[3]{BitVec},
    keywordstyle=[3]\color{teal},
    sensitive=true,
}

\lstdefinelanguage{moxijson}{
    alsoletter={-},
    keywords=[1]{set-logic, define-system, check-system},
    keywordstyle=[1]\color{violet},
    keywords=[2]{input, output, local, init, trans, inv, reachable, query},
    keywordstyle=[2]\color{blue},
    keywords=[3]{BitVec},
    keywordstyle=[3]\color{teal},
    sensitive=true,
}

\newcommand{\mypaperkeywords}{
    Formal verification \and
    Symbolic model checking \and
    Intermediate language \and
    \moxi{} \and
    \btortwo{} \and
    SMT \and
    SAT \and
    \pysmt{} \and
    Exchange formats
}

\ifdefined\techrep
\usepackage[firstpageonly=true,angle=0,vpos=2.8mm,hpos=126.8mm,vanchor=t,hanchor=l]{draftwatermark}
\SetWatermarkText{\href{https://doi.org/10.5281/zenodo.12787654}{\includegraphics[width=0.10\textwidth,alt={artifact available}]{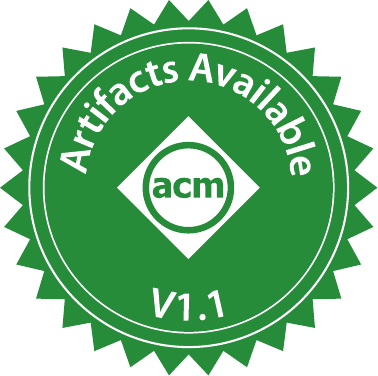}}}
\if

\title{\textsc{MoXIchecker}: \\An Extensible Model Checker for \textsc{MoXI}}

\begin{document}

\ifdefined\techrep
    \author{%
        Salih Ates\orcidID{0009-0001-2331-0955}
        \and
        Dirk Beyer\orcidID{0000-0003-4832-7662}
        \and
        Po-Chun Chien\orcidID{0000-0001-5139-5178}
        \and
        Nian-Ze Lee\orcidID{0000-0002-8096-5595}}

    \institute{LMU Munich, Munich, Germany}
\fi

\maketitle

\begin{abstract}
    \moxi is a new intermediate verification language
introduced in 2024 to promote the standardization and open-source implementations for symbolic model checking
by extending the SMT-LIB~2 language with constructs to define state-transition systems.
The tool suite of \moxi provides a translator from \moxi to \btortwo,
which is a lower-level intermediate language for hardware verification,
and a translation-based model checker,
which invokes mature hardware model checkers for \btortwo to analyze the translated verification tasks.
The extensibility of such a translation-based model checker is restricted
because more complex theories, such as integer or real arithmetics,
cannot be precisely expressed with bit-vectors of fixed lengths in \btortwo.
We present \moxichecker, the first model checker that solves \moxi verification tasks directly.
Instead of translating \moxi to lower-level languages,
\moxichecker uses the solver-agnostic library \pysmt for SMT solvers as backend for its verification algorithms.
\moxichecker is \textit{extensible} because
it accommodates verification tasks involving more complex theories,
not limited by lower-level languages,
facilitates the implementation of new algorithms,
and is solver-agnostic by using the API of \pysmt.
In our evaluation,
\moxichecker uniquely solved tasks that use integer or real arithmetics, and
achieved a comparable performance against
the translation-based model checker from the \moxi tool suite.
    \keywords{\mypaperkeywords}
\end{abstract}

\providecommand\StoreBenchExecResult[7]{\expandafter\newcommand\csname#1#2#3#4#5#6\endcsname{#7}}%
\StoreBenchExecResult{Moxichecker}{KindIncrMsatQFBV}{Status}{All}{}{Score}{0}%
\StoreBenchExecResult{Moxichecker}{KindIncrMsatQFBV}{Status}{All}{}{Count}{382}%
\StoreBenchExecResult{Moxichecker}{KindIncrMsatQFBV}{Status}{Correct}{}{Count}{178}%
\StoreBenchExecResult{Moxichecker}{KindIncrMsatQFBV}{Status}{Correct}{True}{Count}{46}%
\StoreBenchExecResult{Moxichecker}{KindIncrMsatQFBV}{Status}{Correct}{False}{Count}{132}%
\StoreBenchExecResult{Moxichecker}{KindIncrMsatQFBV}{Status}{Wrong}{}{Count}{0}%
\StoreBenchExecResult{Moxichecker}{KindIncrMsatQFBV}{Status}{Wrong}{True}{Count}{0}%
\StoreBenchExecResult{Moxichecker}{KindIncrMsatQFBV}{Status}{Wrong}{False}{Count}{0}%
\StoreBenchExecResult{Moxichecker}{KindIncrMsatQFBV}{Cputime}{All}{}{Sum}{194532.3139179999999999}%
\StoreBenchExecResult{Moxichecker}{KindIncrMsatQFBV}{Cputime}{All}{}{Min}{0.395177}%
\StoreBenchExecResult{Moxichecker}{KindIncrMsatQFBV}{Cputime}{All}{}{Max}{921.519019}%
\StoreBenchExecResult{Moxichecker}{KindIncrMsatQFBV}{Cputime}{All}{}{Avg}{509.2468950732984293191099476}%
\StoreBenchExecResult{Moxichecker}{KindIncrMsatQFBV}{Cputime}{All}{}{Median}{900.228281}%
\StoreBenchExecResult{Moxichecker}{KindIncrMsatQFBV}{Cputime}{All}{}{Stdev}{431.4331210813563503530130119}%
\StoreBenchExecResult{Moxichecker}{KindIncrMsatQFBV}{Cputime}{All}{}{Unit}{s}%
\StoreBenchExecResult{Moxichecker}{KindIncrMsatQFBV}{Cputime}{Correct}{}{Sum}{10582.3131969999999999}%
\StoreBenchExecResult{Moxichecker}{KindIncrMsatQFBV}{Cputime}{Correct}{}{Min}{0.395177}%
\StoreBenchExecResult{Moxichecker}{KindIncrMsatQFBV}{Cputime}{Correct}{}{Max}{841.774753}%
\StoreBenchExecResult{Moxichecker}{KindIncrMsatQFBV}{Cputime}{Correct}{}{Avg}{59.45119773595505617921348315}%
\StoreBenchExecResult{Moxichecker}{KindIncrMsatQFBV}{Cputime}{Correct}{}{Median}{4.6421275}%
\StoreBenchExecResult{Moxichecker}{KindIncrMsatQFBV}{Cputime}{Correct}{}{Stdev}{143.5131450652256013940332360}%
\StoreBenchExecResult{Moxichecker}{KindIncrMsatQFBV}{Cputime}{Correct}{}{Unit}{s}%
\StoreBenchExecResult{Moxichecker}{KindIncrMsatQFBV}{Cputime}{Correct}{True}{Sum}{1154.4048179999999999}%
\StoreBenchExecResult{Moxichecker}{KindIncrMsatQFBV}{Cputime}{Correct}{True}{Min}{0.395177}%
\StoreBenchExecResult{Moxichecker}{KindIncrMsatQFBV}{Cputime}{Correct}{True}{Max}{256.048973}%
\StoreBenchExecResult{Moxichecker}{KindIncrMsatQFBV}{Cputime}{Correct}{True}{Avg}{25.09575691304347825869565217}%
\StoreBenchExecResult{Moxichecker}{KindIncrMsatQFBV}{Cputime}{Correct}{True}{Median}{1.140042}%
\StoreBenchExecResult{Moxichecker}{KindIncrMsatQFBV}{Cputime}{Correct}{True}{Stdev}{56.18540927745213292793386883}%
\StoreBenchExecResult{Moxichecker}{KindIncrMsatQFBV}{Cputime}{Correct}{True}{Unit}{s}%
\StoreBenchExecResult{Moxichecker}{KindIncrMsatQFBV}{Cputime}{Correct}{False}{Sum}{9427.908379}%
\StoreBenchExecResult{Moxichecker}{KindIncrMsatQFBV}{Cputime}{Correct}{False}{Min}{0.427545}%
\StoreBenchExecResult{Moxichecker}{KindIncrMsatQFBV}{Cputime}{Correct}{False}{Max}{841.774753}%
\StoreBenchExecResult{Moxichecker}{KindIncrMsatQFBV}{Cputime}{Correct}{False}{Avg}{71.42354832575757575757575758}%
\StoreBenchExecResult{Moxichecker}{KindIncrMsatQFBV}{Cputime}{Correct}{False}{Median}{6.573026}%
\StoreBenchExecResult{Moxichecker}{KindIncrMsatQFBV}{Cputime}{Correct}{False}{Stdev}{161.6127318249705490160099884}%
\StoreBenchExecResult{Moxichecker}{KindIncrMsatQFBV}{Cputime}{Correct}{False}{Unit}{s}%
\StoreBenchExecResult{Moxichecker}{KindIncrMsatQFBV}{Cputime}{Local}{}{Sum}{194602.607421}%
\StoreBenchExecResult{Moxichecker}{KindIncrMsatQFBV}{Cputime}{Local}{}{Unit}{s}%
\StoreBenchExecResult{Moxichecker}{KindIncrMsatQFBV}{Walltime}{All}{}{Sum}{198994.04053686745411947}%
\StoreBenchExecResult{Moxichecker}{KindIncrMsatQFBV}{Walltime}{All}{}{Min}{0.43478976655751467}%
\StoreBenchExecResult{Moxichecker}{KindIncrMsatQFBV}{Walltime}{All}{}{Max}{973.1556471241638}%
\StoreBenchExecResult{Moxichecker}{KindIncrMsatQFBV}{Walltime}{All}{}{Avg}{520.9268076881346966478272251}%
\StoreBenchExecResult{Moxichecker}{KindIncrMsatQFBV}{Walltime}{All}{}{Median}{900.50215208530425}%
\StoreBenchExecResult{Moxichecker}{KindIncrMsatQFBV}{Walltime}{All}{}{Stdev}{434.9314627562671057542029226}%
\StoreBenchExecResult{Moxichecker}{KindIncrMsatQFBV}{Walltime}{All}{}{Unit}{s}%
\StoreBenchExecResult{Moxichecker}{KindIncrMsatQFBV}{Walltime}{Correct}{}{Sum}{12002.44765450246631947}%
\StoreBenchExecResult{Moxichecker}{KindIncrMsatQFBV}{Walltime}{Correct}{}{Min}{0.43478976655751467}%
\StoreBenchExecResult{Moxichecker}{KindIncrMsatQFBV}{Walltime}{Correct}{}{Max}{843.1992513164878}%
\StoreBenchExecResult{Moxichecker}{KindIncrMsatQFBV}{Walltime}{Correct}{}{Avg}{67.42948120507003550264044944}%
\StoreBenchExecResult{Moxichecker}{KindIncrMsatQFBV}{Walltime}{Correct}{}{Median}{16.1008773078210655}%
\StoreBenchExecResult{Moxichecker}{KindIncrMsatQFBV}{Walltime}{Correct}{}{Stdev}{142.7030624678635102718328182}%
\StoreBenchExecResult{Moxichecker}{KindIncrMsatQFBV}{Walltime}{Correct}{}{Unit}{s}%
\StoreBenchExecResult{Moxichecker}{KindIncrMsatQFBV}{Walltime}{Correct}{True}{Sum}{1510.2417979165911740}%
\StoreBenchExecResult{Moxichecker}{KindIncrMsatQFBV}{Walltime}{Correct}{True}{Min}{0.5198040781542659}%
\StoreBenchExecResult{Moxichecker}{KindIncrMsatQFBV}{Walltime}{Correct}{True}{Max}{259.00398639868945}%
\StoreBenchExecResult{Moxichecker}{KindIncrMsatQFBV}{Walltime}{Correct}{True}{Avg}{32.83134343296937334782608696}%
\StoreBenchExecResult{Moxichecker}{KindIncrMsatQFBV}{Walltime}{Correct}{True}{Median}{16.1008773078210655}%
\StoreBenchExecResult{Moxichecker}{KindIncrMsatQFBV}{Walltime}{Correct}{True}{Stdev}{55.95683620997091402189356511}%
\StoreBenchExecResult{Moxichecker}{KindIncrMsatQFBV}{Walltime}{Correct}{True}{Unit}{s}%
\StoreBenchExecResult{Moxichecker}{KindIncrMsatQFBV}{Walltime}{Correct}{False}{Sum}{10492.20585658587514547}%
\StoreBenchExecResult{Moxichecker}{KindIncrMsatQFBV}{Walltime}{Correct}{False}{Min}{0.43478976655751467}%
\StoreBenchExecResult{Moxichecker}{KindIncrMsatQFBV}{Walltime}{Correct}{False}{Max}{843.1992513164878}%
\StoreBenchExecResult{Moxichecker}{KindIncrMsatQFBV}{Walltime}{Correct}{False}{Avg}{79.48640800443844807174242424}%
\StoreBenchExecResult{Moxichecker}{KindIncrMsatQFBV}{Walltime}{Correct}{False}{Median}{16.0870571685954925}%
\StoreBenchExecResult{Moxichecker}{KindIncrMsatQFBV}{Walltime}{Correct}{False}{Stdev}{160.6458387890566250456386787}%
\StoreBenchExecResult{Moxichecker}{KindIncrMsatQFBV}{Walltime}{Correct}{False}{Unit}{s}%
\StoreBenchExecResult{Moxichecker}{KindIncrMsatQFBV}{Walltime}{Local}{}{Sum}{2764.4014309030026}%
\StoreBenchExecResult{Moxichecker}{KindIncrMsatQFBV}{Walltime}{Local}{}{Unit}{s}%
\providecommand\StoreBenchExecResult[7]{\expandafter\newcommand\csname#1#2#3#4#5#6\endcsname{#7}}%
\StoreBenchExecResult{Moxichecker}{KindMsatQFBV}{Status}{All}{}{Score}{0}%
\StoreBenchExecResult{Moxichecker}{KindMsatQFBV}{Status}{All}{}{Count}{382}%
\StoreBenchExecResult{Moxichecker}{KindMsatQFBV}{Status}{Correct}{}{Count}{169}%
\StoreBenchExecResult{Moxichecker}{KindMsatQFBV}{Status}{Correct}{True}{Count}{44}%
\StoreBenchExecResult{Moxichecker}{KindMsatQFBV}{Status}{Correct}{False}{Count}{125}%
\StoreBenchExecResult{Moxichecker}{KindMsatQFBV}{Status}{Wrong}{}{Count}{0}%
\StoreBenchExecResult{Moxichecker}{KindMsatQFBV}{Status}{Wrong}{True}{Count}{0}%
\StoreBenchExecResult{Moxichecker}{KindMsatQFBV}{Status}{Wrong}{False}{Count}{0}%
\StoreBenchExecResult{Moxichecker}{KindMsatQFBV}{Cputime}{All}{}{Sum}{204631.9968159999999997}%
\StoreBenchExecResult{Moxichecker}{KindMsatQFBV}{Cputime}{All}{}{Min}{0.369828}%
\StoreBenchExecResult{Moxichecker}{KindMsatQFBV}{Cputime}{All}{}{Max}{927.37786}%
\StoreBenchExecResult{Moxichecker}{KindMsatQFBV}{Cputime}{All}{}{Avg}{535.6858555392670157060209424}%
\StoreBenchExecResult{Moxichecker}{KindMsatQFBV}{Cputime}{All}{}{Median}{900.2744455}%
\StoreBenchExecResult{Moxichecker}{KindMsatQFBV}{Cputime}{All}{}{Stdev}{424.3398675355636620829292796}%
\StoreBenchExecResult{Moxichecker}{KindMsatQFBV}{Cputime}{All}{}{Unit}{s}%
\StoreBenchExecResult{Moxichecker}{KindMsatQFBV}{Cputime}{Correct}{}{Sum}{12490.2357599999999997}%
\StoreBenchExecResult{Moxichecker}{KindMsatQFBV}{Cputime}{Correct}{}{Min}{0.369828}%
\StoreBenchExecResult{Moxichecker}{KindMsatQFBV}{Cputime}{Correct}{}{Max}{882.802513}%
\StoreBenchExecResult{Moxichecker}{KindMsatQFBV}{Cputime}{Correct}{}{Avg}{73.90672047337278106331360947}%
\StoreBenchExecResult{Moxichecker}{KindMsatQFBV}{Cputime}{Correct}{}{Median}{5.050645}%
\StoreBenchExecResult{Moxichecker}{KindMsatQFBV}{Cputime}{Correct}{}{Stdev}{156.7158029443199180399652545}%
\StoreBenchExecResult{Moxichecker}{KindMsatQFBV}{Cputime}{Correct}{}{Unit}{s}%
\StoreBenchExecResult{Moxichecker}{KindMsatQFBV}{Cputime}{Correct}{True}{Sum}{1435.8936819999999999}%
\StoreBenchExecResult{Moxichecker}{KindMsatQFBV}{Cputime}{Correct}{True}{Min}{0.369828}%
\StoreBenchExecResult{Moxichecker}{KindMsatQFBV}{Cputime}{Correct}{True}{Max}{349.110707}%
\StoreBenchExecResult{Moxichecker}{KindMsatQFBV}{Cputime}{Correct}{True}{Avg}{32.63394731818181817954545455}%
\StoreBenchExecResult{Moxichecker}{KindMsatQFBV}{Cputime}{Correct}{True}{Median}{1.044523}%
\StoreBenchExecResult{Moxichecker}{KindMsatQFBV}{Cputime}{Correct}{True}{Stdev}{71.39041363057734154792319217}%
\StoreBenchExecResult{Moxichecker}{KindMsatQFBV}{Cputime}{Correct}{True}{Unit}{s}%
\StoreBenchExecResult{Moxichecker}{KindMsatQFBV}{Cputime}{Correct}{False}{Sum}{11054.3420779999999998}%
\StoreBenchExecResult{Moxichecker}{KindMsatQFBV}{Cputime}{Correct}{False}{Min}{0.405499}%
\StoreBenchExecResult{Moxichecker}{KindMsatQFBV}{Cputime}{Correct}{False}{Max}{882.802513}%
\StoreBenchExecResult{Moxichecker}{KindMsatQFBV}{Cputime}{Correct}{False}{Avg}{88.4347366239999999984}%
\StoreBenchExecResult{Moxichecker}{KindMsatQFBV}{Cputime}{Correct}{False}{Median}{6.850324}%
\StoreBenchExecResult{Moxichecker}{KindMsatQFBV}{Cputime}{Correct}{False}{Stdev}{174.9292220718048304752342997}%
\StoreBenchExecResult{Moxichecker}{KindMsatQFBV}{Cputime}{Correct}{False}{Unit}{s}%
\StoreBenchExecResult{Moxichecker}{KindMsatQFBV}{Cputime}{Local}{}{Sum}{204699.987295}%
\StoreBenchExecResult{Moxichecker}{KindMsatQFBV}{Cputime}{Local}{}{Unit}{s}%
\StoreBenchExecResult{Moxichecker}{KindMsatQFBV}{Walltime}{All}{}{Sum}{207028.23696220852412184}%
\StoreBenchExecResult{Moxichecker}{KindMsatQFBV}{Walltime}{All}{}{Min}{0.4009818648919463}%
\StoreBenchExecResult{Moxichecker}{KindMsatQFBV}{Walltime}{All}{}{Max}{939.9584412872791}%
\StoreBenchExecResult{Moxichecker}{KindMsatQFBV}{Walltime}{All}{}{Avg}{541.9587355031636757116230366}%
\StoreBenchExecResult{Moxichecker}{KindMsatQFBV}{Walltime}{All}{}{Median}{900.64690261892975}%
\StoreBenchExecResult{Moxichecker}{KindMsatQFBV}{Walltime}{All}{}{Stdev}{425.1533505418974696889250051}%
\StoreBenchExecResult{Moxichecker}{KindMsatQFBV}{Walltime}{All}{}{Unit}{s}%
\StoreBenchExecResult{Moxichecker}{KindMsatQFBV}{Walltime}{Correct}{}{Sum}{13423.90282614808522184}%
\StoreBenchExecResult{Moxichecker}{KindMsatQFBV}{Walltime}{Correct}{}{Min}{0.4009818648919463}%
\StoreBenchExecResult{Moxichecker}{KindMsatQFBV}{Walltime}{Correct}{}{Max}{885.6308318208903}%
\StoreBenchExecResult{Moxichecker}{KindMsatQFBV}{Walltime}{Correct}{}{Avg}{79.43137766951529717065088757}%
\StoreBenchExecResult{Moxichecker}{KindMsatQFBV}{Walltime}{Correct}{}{Median}{12.868247631005943}%
\StoreBenchExecResult{Moxichecker}{KindMsatQFBV}{Walltime}{Correct}{}{Stdev}{157.4187258161842764784813338}%
\StoreBenchExecResult{Moxichecker}{KindMsatQFBV}{Walltime}{Correct}{}{Unit}{s}%
\StoreBenchExecResult{Moxichecker}{KindMsatQFBV}{Walltime}{Correct}{True}{Sum}{1614.7250690171495097}%
\StoreBenchExecResult{Moxichecker}{KindMsatQFBV}{Walltime}{Correct}{True}{Min}{0.4009818648919463}%
\StoreBenchExecResult{Moxichecker}{KindMsatQFBV}{Walltime}{Correct}{True}{Max}{350.91520637180656}%
\StoreBenchExecResult{Moxichecker}{KindMsatQFBV}{Walltime}{Correct}{True}{Avg}{36.69829702311703431136363636}%
\StoreBenchExecResult{Moxichecker}{KindMsatQFBV}{Walltime}{Correct}{True}{Median}{10.687797470483929}%
\StoreBenchExecResult{Moxichecker}{KindMsatQFBV}{Walltime}{Correct}{True}{Stdev}{70.49193129557808914408842906}%
\StoreBenchExecResult{Moxichecker}{KindMsatQFBV}{Walltime}{Correct}{True}{Unit}{s}%
\StoreBenchExecResult{Moxichecker}{KindMsatQFBV}{Walltime}{Correct}{False}{Sum}{11809.17775713093571214}%
\StoreBenchExecResult{Moxichecker}{KindMsatQFBV}{Walltime}{Correct}{False}{Min}{0.4170906040817499}%
\StoreBenchExecResult{Moxichecker}{KindMsatQFBV}{Walltime}{Correct}{False}{Max}{885.6308318208903}%
\StoreBenchExecResult{Moxichecker}{KindMsatQFBV}{Walltime}{Correct}{False}{Avg}{94.47342205704748569712}%
\StoreBenchExecResult{Moxichecker}{KindMsatQFBV}{Walltime}{Correct}{False}{Median}{16.498100117780268}%
\StoreBenchExecResult{Moxichecker}{KindMsatQFBV}{Walltime}{Correct}{False}{Stdev}{175.7420334723275419426266260}%
\StoreBenchExecResult{Moxichecker}{KindMsatQFBV}{Walltime}{Correct}{False}{Unit}{s}%
\StoreBenchExecResult{Moxichecker}{KindMsatQFBV}{Walltime}{Local}{}{Sum}{2799.0513033997267}%
\StoreBenchExecResult{Moxichecker}{KindMsatQFBV}{Walltime}{Local}{}{Unit}{s}%
\providecommand\StoreBenchExecResult[7]{\expandafter\newcommand\csname#1#2#3#4#5#6\endcsname{#7}}%
\StoreBenchExecResult{Moxichecker}{KindIncrZIIIQFBV}{Status}{All}{}{Score}{0}%
\StoreBenchExecResult{Moxichecker}{KindIncrZIIIQFBV}{Status}{All}{}{Count}{382}%
\StoreBenchExecResult{Moxichecker}{KindIncrZIIIQFBV}{Status}{Correct}{}{Count}{174}%
\StoreBenchExecResult{Moxichecker}{KindIncrZIIIQFBV}{Status}{Correct}{True}{Count}{47}%
\StoreBenchExecResult{Moxichecker}{KindIncrZIIIQFBV}{Status}{Correct}{False}{Count}{127}%
\StoreBenchExecResult{Moxichecker}{KindIncrZIIIQFBV}{Status}{Wrong}{}{Count}{0}%
\StoreBenchExecResult{Moxichecker}{KindIncrZIIIQFBV}{Status}{Wrong}{True}{Count}{0}%
\StoreBenchExecResult{Moxichecker}{KindIncrZIIIQFBV}{Status}{Wrong}{False}{Count}{0}%
\StoreBenchExecResult{Moxichecker}{KindIncrZIIIQFBV}{Cputime}{All}{}{Sum}{196366.1851039999999999}%
\StoreBenchExecResult{Moxichecker}{KindIncrZIIIQFBV}{Cputime}{All}{}{Min}{0.398177}%
\StoreBenchExecResult{Moxichecker}{KindIncrZIIIQFBV}{Cputime}{All}{}{Max}{915.883046}%
\StoreBenchExecResult{Moxichecker}{KindIncrZIIIQFBV}{Cputime}{All}{}{Avg}{514.0476049842931937170157068}%
\StoreBenchExecResult{Moxichecker}{KindIncrZIIIQFBV}{Cputime}{All}{}{Median}{900.2154215}%
\StoreBenchExecResult{Moxichecker}{KindIncrZIIIQFBV}{Cputime}{All}{}{Stdev}{432.9007977038484278614886038}%
\StoreBenchExecResult{Moxichecker}{KindIncrZIIIQFBV}{Cputime}{All}{}{Unit}{s}%
\StoreBenchExecResult{Moxichecker}{KindIncrZIIIQFBV}{Cputime}{Correct}{}{Sum}{8859.9703619999999999}%
\StoreBenchExecResult{Moxichecker}{KindIncrZIIIQFBV}{Cputime}{Correct}{}{Min}{0.398177}%
\StoreBenchExecResult{Moxichecker}{KindIncrZIIIQFBV}{Cputime}{Correct}{}{Max}{739.339716}%
\StoreBenchExecResult{Moxichecker}{KindIncrZIIIQFBV}{Cputime}{Correct}{}{Avg}{50.91936989655172413735632184}%
\StoreBenchExecResult{Moxichecker}{KindIncrZIIIQFBV}{Cputime}{Correct}{}{Median}{3.332046}%
\StoreBenchExecResult{Moxichecker}{KindIncrZIIIQFBV}{Cputime}{Correct}{}{Stdev}{132.2903656902087052510624661}%
\StoreBenchExecResult{Moxichecker}{KindIncrZIIIQFBV}{Cputime}{Correct}{}{Unit}{s}%
\StoreBenchExecResult{Moxichecker}{KindIncrZIIIQFBV}{Cputime}{Correct}{True}{Sum}{1921.038551}%
\StoreBenchExecResult{Moxichecker}{KindIncrZIIIQFBV}{Cputime}{Correct}{True}{Min}{0.398177}%
\StoreBenchExecResult{Moxichecker}{KindIncrZIIIQFBV}{Cputime}{Correct}{True}{Max}{739.339716}%
\StoreBenchExecResult{Moxichecker}{KindIncrZIIIQFBV}{Cputime}{Correct}{True}{Avg}{40.87316065957446808510638298}%
\StoreBenchExecResult{Moxichecker}{KindIncrZIIIQFBV}{Cputime}{Correct}{True}{Median}{1.085834}%
\StoreBenchExecResult{Moxichecker}{KindIncrZIIIQFBV}{Cputime}{Correct}{True}{Stdev}{137.5574366790884015059424099}%
\StoreBenchExecResult{Moxichecker}{KindIncrZIIIQFBV}{Cputime}{Correct}{True}{Unit}{s}%
\StoreBenchExecResult{Moxichecker}{KindIncrZIIIQFBV}{Cputime}{Correct}{False}{Sum}{6938.9318109999999999}%
\StoreBenchExecResult{Moxichecker}{KindIncrZIIIQFBV}{Cputime}{Correct}{False}{Min}{0.432319}%
\StoreBenchExecResult{Moxichecker}{KindIncrZIIIQFBV}{Cputime}{Correct}{False}{Max}{701.031739}%
\StoreBenchExecResult{Moxichecker}{KindIncrZIIIQFBV}{Cputime}{Correct}{False}{Avg}{54.63725835433070866062992126}%
\StoreBenchExecResult{Moxichecker}{KindIncrZIIIQFBV}{Cputime}{Correct}{False}{Median}{4.204534}%
\StoreBenchExecResult{Moxichecker}{KindIncrZIIIQFBV}{Cputime}{Correct}{False}{Stdev}{130.0906290781487959507676867}%
\StoreBenchExecResult{Moxichecker}{KindIncrZIIIQFBV}{Cputime}{Correct}{False}{Unit}{s}%
\StoreBenchExecResult{Moxichecker}{KindIncrZIIIQFBV}{Cputime}{Local}{}{Sum}{196433.630792}%
\StoreBenchExecResult{Moxichecker}{KindIncrZIIIQFBV}{Cputime}{Local}{}{Unit}{s}%
\StoreBenchExecResult{Moxichecker}{KindIncrZIIIQFBV}{Walltime}{All}{}{Sum}{198788.06110483314857623}%
\StoreBenchExecResult{Moxichecker}{KindIncrZIIIQFBV}{Walltime}{All}{}{Min}{0.4332708930596709}%
\StoreBenchExecResult{Moxichecker}{KindIncrZIIIQFBV}{Walltime}{All}{}{Max}{941.1242055473849}%
\StoreBenchExecResult{Moxichecker}{KindIncrZIIIQFBV}{Walltime}{All}{}{Avg}{520.3875945152700224508638743}%
\StoreBenchExecResult{Moxichecker}{KindIncrZIIIQFBV}{Walltime}{All}{}{Median}{900.4915404776111}%
\StoreBenchExecResult{Moxichecker}{KindIncrZIIIQFBV}{Walltime}{All}{}{Stdev}{433.9349602262613726188003365}%
\StoreBenchExecResult{Moxichecker}{KindIncrZIIIQFBV}{Walltime}{All}{}{Unit}{s}%
\StoreBenchExecResult{Moxichecker}{KindIncrZIIIQFBV}{Walltime}{Correct}{}{Sum}{9786.53422608412797623}%
\StoreBenchExecResult{Moxichecker}{KindIncrZIIIQFBV}{Walltime}{Correct}{}{Min}{0.4332708930596709}%
\StoreBenchExecResult{Moxichecker}{KindIncrZIIIQFBV}{Walltime}{Correct}{}{Max}{743.2312166430056}%
\StoreBenchExecResult{Moxichecker}{KindIncrZIIIQFBV}{Walltime}{Correct}{}{Avg}{56.24444957519613779442528736}%
\StoreBenchExecResult{Moxichecker}{KindIncrZIIIQFBV}{Walltime}{Correct}{}{Median}{10.352240427862853}%
\StoreBenchExecResult{Moxichecker}{KindIncrZIIIQFBV}{Walltime}{Correct}{}{Stdev}{132.8360765023016152599547083}%
\StoreBenchExecResult{Moxichecker}{KindIncrZIIIQFBV}{Walltime}{Correct}{}{Unit}{s}%
\StoreBenchExecResult{Moxichecker}{KindIncrZIIIQFBV}{Walltime}{Correct}{True}{Sum}{2122.10370874684306383}%
\StoreBenchExecResult{Moxichecker}{KindIncrZIIIQFBV}{Walltime}{Correct}{True}{Min}{0.4332708930596709}%
\StoreBenchExecResult{Moxichecker}{KindIncrZIIIQFBV}{Walltime}{Correct}{True}{Max}{743.2312166430056}%
\StoreBenchExecResult{Moxichecker}{KindIncrZIIIQFBV}{Walltime}{Correct}{True}{Avg}{45.15114273929453327297872340}%
\StoreBenchExecResult{Moxichecker}{KindIncrZIIIQFBV}{Walltime}{Correct}{True}{Median}{9.292078420519829}%
\StoreBenchExecResult{Moxichecker}{KindIncrZIIIQFBV}{Walltime}{Correct}{True}{Stdev}{137.5333053050379495329115079}%
\StoreBenchExecResult{Moxichecker}{KindIncrZIIIQFBV}{Walltime}{Correct}{True}{Unit}{s}%
\StoreBenchExecResult{Moxichecker}{KindIncrZIIIQFBV}{Walltime}{Correct}{False}{Sum}{7664.4305173372849124}%
\StoreBenchExecResult{Moxichecker}{KindIncrZIIIQFBV}{Walltime}{Correct}{False}{Min}{0.4576269155368209}%
\StoreBenchExecResult{Moxichecker}{KindIncrZIIIQFBV}{Walltime}{Correct}{False}{Max}{703.9506808668375}%
\StoreBenchExecResult{Moxichecker}{KindIncrZIIIQFBV}{Walltime}{Correct}{False}{Avg}{60.34984659320696781417322835}%
\StoreBenchExecResult{Moxichecker}{KindIncrZIIIQFBV}{Walltime}{Correct}{False}{Median}{11.024967296980321}%
\StoreBenchExecResult{Moxichecker}{KindIncrZIIIQFBV}{Walltime}{Correct}{False}{Stdev}{130.8167855226086415614384392}%
\StoreBenchExecResult{Moxichecker}{KindIncrZIIIQFBV}{Walltime}{Correct}{False}{Unit}{s}%
\StoreBenchExecResult{Moxichecker}{KindIncrZIIIQFBV}{Walltime}{Local}{}{Sum}{2770.8582982458174}%
\StoreBenchExecResult{Moxichecker}{KindIncrZIIIQFBV}{Walltime}{Local}{}{Unit}{s}%
\providecommand\StoreBenchExecResult[7]{\expandafter\newcommand\csname#1#2#3#4#5#6\endcsname{#7}}%
\StoreBenchExecResult{Moxichecker}{KindZIIIQFBV}{Status}{All}{}{Score}{0}%
\StoreBenchExecResult{Moxichecker}{KindZIIIQFBV}{Status}{All}{}{Count}{382}%
\StoreBenchExecResult{Moxichecker}{KindZIIIQFBV}{Status}{Correct}{}{Count}{169}%
\StoreBenchExecResult{Moxichecker}{KindZIIIQFBV}{Status}{Correct}{True}{Count}{47}%
\StoreBenchExecResult{Moxichecker}{KindZIIIQFBV}{Status}{Correct}{False}{Count}{122}%
\StoreBenchExecResult{Moxichecker}{KindZIIIQFBV}{Status}{Wrong}{}{Count}{0}%
\StoreBenchExecResult{Moxichecker}{KindZIIIQFBV}{Status}{Wrong}{True}{Count}{0}%
\StoreBenchExecResult{Moxichecker}{KindZIIIQFBV}{Status}{Wrong}{False}{Count}{0}%
\StoreBenchExecResult{Moxichecker}{KindZIIIQFBV}{Cputime}{All}{}{Sum}{204435.9632729999999998}%
\StoreBenchExecResult{Moxichecker}{KindZIIIQFBV}{Cputime}{All}{}{Min}{0.398247}%
\StoreBenchExecResult{Moxichecker}{KindZIIIQFBV}{Cputime}{All}{}{Max}{915.954521}%
\StoreBenchExecResult{Moxichecker}{KindZIIIQFBV}{Cputime}{All}{}{Avg}{535.1726787251308900518324607}%
\StoreBenchExecResult{Moxichecker}{KindZIIIQFBV}{Cputime}{All}{}{Median}{900.220917}%
\StoreBenchExecResult{Moxichecker}{KindZIIIQFBV}{Cputime}{All}{}{Stdev}{425.2645698275506808718594076}%
\StoreBenchExecResult{Moxichecker}{KindZIIIQFBV}{Cputime}{All}{}{Unit}{s}%
\StoreBenchExecResult{Moxichecker}{KindZIIIQFBV}{Cputime}{Correct}{}{Sum}{12427.4194349999999998}%
\StoreBenchExecResult{Moxichecker}{KindZIIIQFBV}{Cputime}{Correct}{}{Min}{0.398247}%
\StoreBenchExecResult{Moxichecker}{KindZIIIQFBV}{Cputime}{Correct}{}{Max}{860.443562}%
\StoreBenchExecResult{Moxichecker}{KindZIIIQFBV}{Cputime}{Correct}{}{Avg}{73.53502624260355029467455621}%
\StoreBenchExecResult{Moxichecker}{KindZIIIQFBV}{Cputime}{Correct}{}{Median}{4.10552}%
\StoreBenchExecResult{Moxichecker}{KindZIIIQFBV}{Cputime}{Correct}{}{Stdev}{163.0341979940934711619141591}%
\StoreBenchExecResult{Moxichecker}{KindZIIIQFBV}{Cputime}{Correct}{}{Unit}{s}%
\StoreBenchExecResult{Moxichecker}{KindZIIIQFBV}{Cputime}{Correct}{True}{Sum}{2668.251222}%
\StoreBenchExecResult{Moxichecker}{KindZIIIQFBV}{Cputime}{Correct}{True}{Min}{0.398247}%
\StoreBenchExecResult{Moxichecker}{KindZIIIQFBV}{Cputime}{Correct}{True}{Max}{828.875063}%
\StoreBenchExecResult{Moxichecker}{KindZIIIQFBV}{Cputime}{Correct}{True}{Avg}{56.77130259574468085106382979}%
\StoreBenchExecResult{Moxichecker}{KindZIIIQFBV}{Cputime}{Correct}{True}{Median}{1.099285}%
\StoreBenchExecResult{Moxichecker}{KindZIIIQFBV}{Cputime}{Correct}{True}{Stdev}{155.1024522143380333083193080}%
\StoreBenchExecResult{Moxichecker}{KindZIIIQFBV}{Cputime}{Correct}{True}{Unit}{s}%
\StoreBenchExecResult{Moxichecker}{KindZIIIQFBV}{Cputime}{Correct}{False}{Sum}{9759.1682129999999998}%
\StoreBenchExecResult{Moxichecker}{KindZIIIQFBV}{Cputime}{Correct}{False}{Min}{0.42811}%
\StoreBenchExecResult{Moxichecker}{KindZIIIQFBV}{Cputime}{Correct}{False}{Max}{860.443562}%
\StoreBenchExecResult{Moxichecker}{KindZIIIQFBV}{Cputime}{Correct}{False}{Avg}{79.99318207377049180163934426}%
\StoreBenchExecResult{Moxichecker}{KindZIIIQFBV}{Cputime}{Correct}{False}{Median}{6.1656645}%
\StoreBenchExecResult{Moxichecker}{KindZIIIQFBV}{Cputime}{Correct}{False}{Stdev}{165.5364027331655677702309725}%
\StoreBenchExecResult{Moxichecker}{KindZIIIQFBV}{Cputime}{Correct}{False}{Unit}{s}%
\StoreBenchExecResult{Moxichecker}{KindZIIIQFBV}{Cputime}{Local}{}{Sum}{204502.833667}%
\StoreBenchExecResult{Moxichecker}{KindZIIIQFBV}{Cputime}{Local}{}{Unit}{s}%
\StoreBenchExecResult{Moxichecker}{KindZIIIQFBV}{Walltime}{All}{}{Sum}{207037.58676692564005795}%
\StoreBenchExecResult{Moxichecker}{KindZIIIQFBV}{Walltime}{All}{}{Min}{0.41584166418761015}%
\StoreBenchExecResult{Moxichecker}{KindZIIIQFBV}{Walltime}{All}{}{Max}{948.2189552467316}%
\StoreBenchExecResult{Moxichecker}{KindZIIIQFBV}{Walltime}{All}{}{Avg}{541.9832114317425132407068063}%
\StoreBenchExecResult{Moxichecker}{KindZIIIQFBV}{Walltime}{All}{}{Median}{900.63870945526285}%
\StoreBenchExecResult{Moxichecker}{KindZIIIQFBV}{Walltime}{All}{}{Stdev}{426.5163434730615915169803253}%
\StoreBenchExecResult{Moxichecker}{KindZIIIQFBV}{Walltime}{All}{}{Unit}{s}%
\StoreBenchExecResult{Moxichecker}{KindZIIIQFBV}{Walltime}{Correct}{}{Sum}{13414.70384723972525795}%
\StoreBenchExecResult{Moxichecker}{KindZIIIQFBV}{Walltime}{Correct}{}{Min}{0.41584166418761015}%
\StoreBenchExecResult{Moxichecker}{KindZIIIQFBV}{Walltime}{Correct}{}{Max}{861.3529620729387}%
\StoreBenchExecResult{Moxichecker}{KindZIIIQFBV}{Walltime}{Correct}{}{Avg}{79.37694584165517904112426036}%
\StoreBenchExecResult{Moxichecker}{KindZIIIQFBV}{Walltime}{Correct}{}{Median}{11.10449746903032}%
\StoreBenchExecResult{Moxichecker}{KindZIIIQFBV}{Walltime}{Correct}{}{Stdev}{165.0662239375547160438855365}%
\StoreBenchExecResult{Moxichecker}{KindZIIIQFBV}{Walltime}{Correct}{}{Unit}{s}%
\StoreBenchExecResult{Moxichecker}{KindZIIIQFBV}{Walltime}{Correct}{True}{Sum}{2876.49134558904912675}%
\StoreBenchExecResult{Moxichecker}{KindZIIIQFBV}{Walltime}{Correct}{True}{Min}{0.41584166418761015}%
\StoreBenchExecResult{Moxichecker}{KindZIIIQFBV}{Walltime}{Correct}{True}{Max}{829.1314860396087}%
\StoreBenchExecResult{Moxichecker}{KindZIIIQFBV}{Walltime}{Correct}{True}{Avg}{61.20194352317125801595744681}%
\StoreBenchExecResult{Moxichecker}{KindZIIIQFBV}{Walltime}{Correct}{True}{Median}{10.423981548286974}%
\StoreBenchExecResult{Moxichecker}{KindZIIIQFBV}{Walltime}{Correct}{True}{Stdev}{153.8650636381614860764265571}%
\StoreBenchExecResult{Moxichecker}{KindZIIIQFBV}{Walltime}{Correct}{True}{Unit}{s}%
\StoreBenchExecResult{Moxichecker}{KindZIIIQFBV}{Walltime}{Correct}{False}{Sum}{10538.2125016506761312}%
\StoreBenchExecResult{Moxichecker}{KindZIIIQFBV}{Walltime}{Correct}{False}{Min}{0.4766789944842458}%
\StoreBenchExecResult{Moxichecker}{KindZIIIQFBV}{Walltime}{Correct}{False}{Max}{861.3529620729387}%
\StoreBenchExecResult{Moxichecker}{KindZIIIQFBV}{Walltime}{Correct}{False}{Avg}{86.3787909971366896}%
\StoreBenchExecResult{Moxichecker}{KindZIIIQFBV}{Walltime}{Correct}{False}{Median}{11.10666884155944}%
\StoreBenchExecResult{Moxichecker}{KindZIIIQFBV}{Walltime}{Correct}{False}{Stdev}{168.6618704394157238766940619}%
\StoreBenchExecResult{Moxichecker}{KindZIIIQFBV}{Walltime}{Correct}{False}{Unit}{s}%
\StoreBenchExecResult{Moxichecker}{KindZIIIQFBV}{Walltime}{Local}{}{Sum}{2781.81423226092}%
\StoreBenchExecResult{Moxichecker}{KindZIIIQFBV}{Walltime}{Local}{}{Unit}{s}%
\providecommand\StoreBenchExecResult[7]{\expandafter\newcommand\csname#1#2#3#4#5#6\endcsname{#7}}%
\StoreBenchExecResult{MoxiMcFlow}{AvrKindQFBV}{Status}{All}{}{Score}{0}%
\StoreBenchExecResult{MoxiMcFlow}{AvrKindQFBV}{Status}{All}{}{Count}{382}%
\StoreBenchExecResult{MoxiMcFlow}{AvrKindQFBV}{Status}{Correct}{}{Count}{175}%
\StoreBenchExecResult{MoxiMcFlow}{AvrKindQFBV}{Status}{Correct}{True}{Count}{45}%
\StoreBenchExecResult{MoxiMcFlow}{AvrKindQFBV}{Status}{Correct}{False}{Count}{130}%
\StoreBenchExecResult{MoxiMcFlow}{AvrKindQFBV}{Status}{Wrong}{}{Count}{0}%
\StoreBenchExecResult{MoxiMcFlow}{AvrKindQFBV}{Status}{Wrong}{True}{Count}{0}%
\StoreBenchExecResult{MoxiMcFlow}{AvrKindQFBV}{Status}{Wrong}{False}{Count}{0}%
\StoreBenchExecResult{MoxiMcFlow}{AvrKindQFBV}{Cputime}{All}{}{Sum}{145005.994547}%
\StoreBenchExecResult{MoxiMcFlow}{AvrKindQFBV}{Cputime}{All}{}{Min}{2.822207}%
\StoreBenchExecResult{MoxiMcFlow}{AvrKindQFBV}{Cputime}{All}{}{Max}{913.256508}%
\StoreBenchExecResult{MoxiMcFlow}{AvrKindQFBV}{Cputime}{All}{}{Avg}{379.5968443638743455497382199}%
\StoreBenchExecResult{MoxiMcFlow}{AvrKindQFBV}{Cputime}{All}{}{Median}{44.822792}%
\StoreBenchExecResult{MoxiMcFlow}{AvrKindQFBV}{Cputime}{All}{}{Stdev}{426.5999126403182097998557838}%
\StoreBenchExecResult{MoxiMcFlow}{AvrKindQFBV}{Cputime}{All}{}{Unit}{s}%
\StoreBenchExecResult{MoxiMcFlow}{AvrKindQFBV}{Cputime}{Correct}{}{Sum}{9361.282179}%
\StoreBenchExecResult{MoxiMcFlow}{AvrKindQFBV}{Cputime}{Correct}{}{Min}{2.822207}%
\StoreBenchExecResult{MoxiMcFlow}{AvrKindQFBV}{Cputime}{Correct}{}{Max}{728.7236}%
\StoreBenchExecResult{MoxiMcFlow}{AvrKindQFBV}{Cputime}{Correct}{}{Avg}{53.49304102285714285714285714}%
\StoreBenchExecResult{MoxiMcFlow}{AvrKindQFBV}{Cputime}{Correct}{}{Median}{11.180329}%
\StoreBenchExecResult{MoxiMcFlow}{AvrKindQFBV}{Cputime}{Correct}{}{Stdev}{110.7108740210881616826996747}%
\StoreBenchExecResult{MoxiMcFlow}{AvrKindQFBV}{Cputime}{Correct}{}{Unit}{s}%
\StoreBenchExecResult{MoxiMcFlow}{AvrKindQFBV}{Cputime}{Correct}{True}{Sum}{1417.132681}%
\StoreBenchExecResult{MoxiMcFlow}{AvrKindQFBV}{Cputime}{Correct}{True}{Min}{2.822207}%
\StoreBenchExecResult{MoxiMcFlow}{AvrKindQFBV}{Cputime}{Correct}{True}{Max}{271.553355}%
\StoreBenchExecResult{MoxiMcFlow}{AvrKindQFBV}{Cputime}{Correct}{True}{Avg}{31.49183735555555555555555556}%
\StoreBenchExecResult{MoxiMcFlow}{AvrKindQFBV}{Cputime}{Correct}{True}{Median}{5.737004}%
\StoreBenchExecResult{MoxiMcFlow}{AvrKindQFBV}{Cputime}{Correct}{True}{Stdev}{64.40618327948895524103773400}%
\StoreBenchExecResult{MoxiMcFlow}{AvrKindQFBV}{Cputime}{Correct}{True}{Unit}{s}%
\StoreBenchExecResult{MoxiMcFlow}{AvrKindQFBV}{Cputime}{Correct}{False}{Sum}{7944.149498}%
\StoreBenchExecResult{MoxiMcFlow}{AvrKindQFBV}{Cputime}{Correct}{False}{Min}{3.062331}%
\StoreBenchExecResult{MoxiMcFlow}{AvrKindQFBV}{Cputime}{Correct}{False}{Max}{728.7236}%
\StoreBenchExecResult{MoxiMcFlow}{AvrKindQFBV}{Cputime}{Correct}{False}{Avg}{61.10884229230769230769230769}%
\StoreBenchExecResult{MoxiMcFlow}{AvrKindQFBV}{Cputime}{Correct}{False}{Median}{18.616162}%
\StoreBenchExecResult{MoxiMcFlow}{AvrKindQFBV}{Cputime}{Correct}{False}{Stdev}{121.8122011552750708242323362}%
\StoreBenchExecResult{MoxiMcFlow}{AvrKindQFBV}{Cputime}{Correct}{False}{Unit}{s}%
\StoreBenchExecResult{MoxiMcFlow}{AvrKindQFBV}{Cputime}{Local}{}{Sum}{145074.56032699998}%
\StoreBenchExecResult{MoxiMcFlow}{AvrKindQFBV}{Cputime}{Local}{}{Unit}{s}%
\StoreBenchExecResult{MoxiMcFlow}{AvrKindQFBV}{Walltime}{All}{}{Sum}{146924.4135711835696881}%
\StoreBenchExecResult{MoxiMcFlow}{AvrKindQFBV}{Walltime}{All}{}{Min}{2.8301253747195005}%
\StoreBenchExecResult{MoxiMcFlow}{AvrKindQFBV}{Walltime}{All}{}{Max}{943.7939889850095}%
\StoreBenchExecResult{MoxiMcFlow}{AvrKindQFBV}{Walltime}{All}{}{Avg}{384.6188836941978264086387435}%
\StoreBenchExecResult{MoxiMcFlow}{AvrKindQFBV}{Walltime}{All}{}{Median}{51.948850249871612}%
\StoreBenchExecResult{MoxiMcFlow}{AvrKindQFBV}{Walltime}{All}{}{Stdev}{427.5559605115019711835750428}%
\StoreBenchExecResult{MoxiMcFlow}{AvrKindQFBV}{Walltime}{All}{}{Unit}{s}%
\StoreBenchExecResult{MoxiMcFlow}{AvrKindQFBV}{Walltime}{Correct}{}{Sum}{10175.5172552838921206}%
\StoreBenchExecResult{MoxiMcFlow}{AvrKindQFBV}{Walltime}{Correct}{}{Min}{2.8301253747195005}%
\StoreBenchExecResult{MoxiMcFlow}{AvrKindQFBV}{Walltime}{Correct}{}{Max}{729.9008638849482}%
\StoreBenchExecResult{MoxiMcFlow}{AvrKindQFBV}{Walltime}{Correct}{}{Avg}{58.14581288733652640342857143}%
\StoreBenchExecResult{MoxiMcFlow}{AvrKindQFBV}{Walltime}{Correct}{}{Median}{20.034227535128593}%
\StoreBenchExecResult{MoxiMcFlow}{AvrKindQFBV}{Walltime}{Correct}{}{Stdev}{112.0126487138437235581746569}%
\StoreBenchExecResult{MoxiMcFlow}{AvrKindQFBV}{Walltime}{Correct}{}{Unit}{s}%
\StoreBenchExecResult{MoxiMcFlow}{AvrKindQFBV}{Walltime}{Correct}{True}{Sum}{1608.2396098747849242}%
\StoreBenchExecResult{MoxiMcFlow}{AvrKindQFBV}{Walltime}{Correct}{True}{Min}{2.8301253747195005}%
\StoreBenchExecResult{MoxiMcFlow}{AvrKindQFBV}{Walltime}{Correct}{True}{Max}{281.61300414893776}%
\StoreBenchExecResult{MoxiMcFlow}{AvrKindQFBV}{Walltime}{Correct}{True}{Avg}{35.73865799721744276}%
\StoreBenchExecResult{MoxiMcFlow}{AvrKindQFBV}{Walltime}{Correct}{True}{Median}{12.257859812118113}%
\StoreBenchExecResult{MoxiMcFlow}{AvrKindQFBV}{Walltime}{Correct}{True}{Stdev}{64.92616732521533273239287855}%
\StoreBenchExecResult{MoxiMcFlow}{AvrKindQFBV}{Walltime}{Correct}{True}{Unit}{s}%
\StoreBenchExecResult{MoxiMcFlow}{AvrKindQFBV}{Walltime}{Correct}{False}{Sum}{8567.2776454091071964}%
\StoreBenchExecResult{MoxiMcFlow}{AvrKindQFBV}{Walltime}{Correct}{False}{Min}{3.0801879381760955}%
\StoreBenchExecResult{MoxiMcFlow}{AvrKindQFBV}{Walltime}{Correct}{False}{Max}{729.9008638849482}%
\StoreBenchExecResult{MoxiMcFlow}{AvrKindQFBV}{Walltime}{Correct}{False}{Avg}{65.90213573391620920307692308}%
\StoreBenchExecResult{MoxiMcFlow}{AvrKindQFBV}{Walltime}{Correct}{False}{Median}{25.777673612814397}%
\StoreBenchExecResult{MoxiMcFlow}{AvrKindQFBV}{Walltime}{Correct}{False}{Stdev}{123.2754275884927829313885087}%
\StoreBenchExecResult{MoxiMcFlow}{AvrKindQFBV}{Walltime}{Correct}{False}{Unit}{s}%
\StoreBenchExecResult{MoxiMcFlow}{AvrKindQFBV}{Walltime}{Local}{}{Sum}{2007.7425677431747}%
\StoreBenchExecResult{MoxiMcFlow}{AvrKindQFBV}{Walltime}{Local}{}{Unit}{s}%
\providecommand\StoreBenchExecResult[7]{\expandafter\newcommand\csname#1#2#3#4#5#6\endcsname{#7}}%
\StoreBenchExecResult{MoxiMcFlow}{PonoKindQFBV}{Status}{All}{}{Score}{0}%
\StoreBenchExecResult{MoxiMcFlow}{PonoKindQFBV}{Status}{All}{}{Count}{382}%
\StoreBenchExecResult{MoxiMcFlow}{PonoKindQFBV}{Status}{Correct}{}{Count}{172}%
\StoreBenchExecResult{MoxiMcFlow}{PonoKindQFBV}{Status}{Correct}{True}{Count}{45}%
\StoreBenchExecResult{MoxiMcFlow}{PonoKindQFBV}{Status}{Correct}{False}{Count}{127}%
\StoreBenchExecResult{MoxiMcFlow}{PonoKindQFBV}{Status}{Wrong}{}{Count}{0}%
\StoreBenchExecResult{MoxiMcFlow}{PonoKindQFBV}{Status}{Wrong}{True}{Count}{0}%
\StoreBenchExecResult{MoxiMcFlow}{PonoKindQFBV}{Status}{Wrong}{False}{Count}{0}%
\StoreBenchExecResult{MoxiMcFlow}{PonoKindQFBV}{Cputime}{All}{}{Sum}{146916.440783}%
\StoreBenchExecResult{MoxiMcFlow}{PonoKindQFBV}{Cputime}{All}{}{Min}{2.697972}%
\StoreBenchExecResult{MoxiMcFlow}{PonoKindQFBV}{Cputime}{All}{}{Max}{916.774117}%
\StoreBenchExecResult{MoxiMcFlow}{PonoKindQFBV}{Cputime}{All}{}{Avg}{384.5980125209424083769633508}%
\StoreBenchExecResult{MoxiMcFlow}{PonoKindQFBV}{Cputime}{All}{}{Median}{42.9944205}%
\StoreBenchExecResult{MoxiMcFlow}{PonoKindQFBV}{Cputime}{All}{}{Stdev}{429.2051597265113585503720638}%
\StoreBenchExecResult{MoxiMcFlow}{PonoKindQFBV}{Cputime}{All}{}{Unit}{s}%
\StoreBenchExecResult{MoxiMcFlow}{PonoKindQFBV}{Cputime}{Correct}{}{Sum}{8481.919809}%
\StoreBenchExecResult{MoxiMcFlow}{PonoKindQFBV}{Cputime}{Correct}{}{Min}{2.697972}%
\StoreBenchExecResult{MoxiMcFlow}{PonoKindQFBV}{Cputime}{Correct}{}{Max}{847.469142}%
\StoreBenchExecResult{MoxiMcFlow}{PonoKindQFBV}{Cputime}{Correct}{}{Avg}{49.31348726162790697674418605}%
\StoreBenchExecResult{MoxiMcFlow}{PonoKindQFBV}{Cputime}{Correct}{}{Median}{9.747564}%
\StoreBenchExecResult{MoxiMcFlow}{PonoKindQFBV}{Cputime}{Correct}{}{Stdev}{106.2887189688203653376185094}%
\StoreBenchExecResult{MoxiMcFlow}{PonoKindQFBV}{Cputime}{Correct}{}{Unit}{s}%
\StoreBenchExecResult{MoxiMcFlow}{PonoKindQFBV}{Cputime}{Correct}{True}{Sum}{1287.059636}%
\StoreBenchExecResult{MoxiMcFlow}{PonoKindQFBV}{Cputime}{Correct}{True}{Min}{2.697972}%
\StoreBenchExecResult{MoxiMcFlow}{PonoKindQFBV}{Cputime}{Correct}{True}{Max}{271.738212}%
\StoreBenchExecResult{MoxiMcFlow}{PonoKindQFBV}{Cputime}{Correct}{True}{Avg}{28.60132524444444444444444444}%
\StoreBenchExecResult{MoxiMcFlow}{PonoKindQFBV}{Cputime}{Correct}{True}{Median}{4.589284}%
\StoreBenchExecResult{MoxiMcFlow}{PonoKindQFBV}{Cputime}{Correct}{True}{Stdev}{61.32069995715095051797612377}%
\StoreBenchExecResult{MoxiMcFlow}{PonoKindQFBV}{Cputime}{Correct}{True}{Unit}{s}%
\StoreBenchExecResult{MoxiMcFlow}{PonoKindQFBV}{Cputime}{Correct}{False}{Sum}{7194.860173}%
\StoreBenchExecResult{MoxiMcFlow}{PonoKindQFBV}{Cputime}{Correct}{False}{Min}{2.778888}%
\StoreBenchExecResult{MoxiMcFlow}{PonoKindQFBV}{Cputime}{Correct}{False}{Max}{847.469142}%
\StoreBenchExecResult{MoxiMcFlow}{PonoKindQFBV}{Cputime}{Correct}{False}{Avg}{56.65244230708661417322834646}%
\StoreBenchExecResult{MoxiMcFlow}{PonoKindQFBV}{Cputime}{Correct}{False}{Median}{12.995538}%
\StoreBenchExecResult{MoxiMcFlow}{PonoKindQFBV}{Cputime}{Correct}{False}{Stdev}{117.3117170497260543697804520}%
\StoreBenchExecResult{MoxiMcFlow}{PonoKindQFBV}{Cputime}{Correct}{False}{Unit}{s}%
\StoreBenchExecResult{MoxiMcFlow}{PonoKindQFBV}{Cputime}{Local}{}{Sum}{146984.10666900003}%
\StoreBenchExecResult{MoxiMcFlow}{PonoKindQFBV}{Cputime}{Local}{}{Unit}{s}%
\StoreBenchExecResult{MoxiMcFlow}{PonoKindQFBV}{Walltime}{All}{}{Sum}{149099.6869204621758742}%
\StoreBenchExecResult{MoxiMcFlow}{PonoKindQFBV}{Walltime}{All}{}{Min}{2.7221597246825695}%
\StoreBenchExecResult{MoxiMcFlow}{PonoKindQFBV}{Walltime}{All}{}{Max}{939.8707372108474}%
\StoreBenchExecResult{MoxiMcFlow}{PonoKindQFBV}{Walltime}{All}{}{Avg}{390.3133165457125022884816754}%
\StoreBenchExecResult{MoxiMcFlow}{PonoKindQFBV}{Walltime}{All}{}{Median}{58.0227678427472725}%
\StoreBenchExecResult{MoxiMcFlow}{PonoKindQFBV}{Walltime}{All}{}{Stdev}{429.8713263897006080673051266}%
\StoreBenchExecResult{MoxiMcFlow}{PonoKindQFBV}{Walltime}{All}{}{Unit}{s}%
\StoreBenchExecResult{MoxiMcFlow}{PonoKindQFBV}{Walltime}{Correct}{}{Sum}{9523.8481321148574056}%
\StoreBenchExecResult{MoxiMcFlow}{PonoKindQFBV}{Walltime}{Correct}{}{Min}{2.7221597246825695}%
\StoreBenchExecResult{MoxiMcFlow}{PonoKindQFBV}{Walltime}{Correct}{}{Max}{851.0780446846038}%
\StoreBenchExecResult{MoxiMcFlow}{PonoKindQFBV}{Walltime}{Correct}{}{Avg}{55.37121007043521747441860465}%
\StoreBenchExecResult{MoxiMcFlow}{PonoKindQFBV}{Walltime}{Correct}{}{Median}{17.9517530100420115}%
\StoreBenchExecResult{MoxiMcFlow}{PonoKindQFBV}{Walltime}{Correct}{}{Stdev}{106.6392655059803045661741839}%
\StoreBenchExecResult{MoxiMcFlow}{PonoKindQFBV}{Walltime}{Correct}{}{Unit}{s}%
\StoreBenchExecResult{MoxiMcFlow}{PonoKindQFBV}{Walltime}{Correct}{True}{Sum}{1458.1567894127219886}%
\StoreBenchExecResult{MoxiMcFlow}{PonoKindQFBV}{Walltime}{Correct}{True}{Min}{2.7221597246825695}%
\StoreBenchExecResult{MoxiMcFlow}{PonoKindQFBV}{Walltime}{Correct}{True}{Max}{272.92099368944764}%
\StoreBenchExecResult{MoxiMcFlow}{PonoKindQFBV}{Walltime}{Correct}{True}{Avg}{32.40348420917159974666666667}%
\StoreBenchExecResult{MoxiMcFlow}{PonoKindQFBV}{Walltime}{Correct}{True}{Median}{9.714053146541119}%
\StoreBenchExecResult{MoxiMcFlow}{PonoKindQFBV}{Walltime}{Correct}{True}{Stdev}{60.83849654490160762366845952}%
\StoreBenchExecResult{MoxiMcFlow}{PonoKindQFBV}{Walltime}{Correct}{True}{Unit}{s}%
\StoreBenchExecResult{MoxiMcFlow}{PonoKindQFBV}{Walltime}{Correct}{False}{Sum}{8065.6913427021354170}%
\StoreBenchExecResult{MoxiMcFlow}{PonoKindQFBV}{Walltime}{Correct}{False}{Min}{2.7935156729072332}%
\StoreBenchExecResult{MoxiMcFlow}{PonoKindQFBV}{Walltime}{Correct}{False}{Max}{851.0780446846038}%
\StoreBenchExecResult{MoxiMcFlow}{PonoKindQFBV}{Walltime}{Correct}{False}{Avg}{63.50938065119791666929133858}%
\StoreBenchExecResult{MoxiMcFlow}{PonoKindQFBV}{Walltime}{Correct}{False}{Median}{25.825430774129927}%
\StoreBenchExecResult{MoxiMcFlow}{PonoKindQFBV}{Walltime}{Correct}{False}{Stdev}{117.6295908127677031620905522}%
\StoreBenchExecResult{MoxiMcFlow}{PonoKindQFBV}{Walltime}{Correct}{False}{Unit}{s}%
\StoreBenchExecResult{MoxiMcFlow}{PonoKindQFBV}{Walltime}{Local}{}{Sum}{2010.6192978387699}%
\StoreBenchExecResult{MoxiMcFlow}{PonoKindQFBV}{Walltime}{Local}{}{Unit}{s}%
\edef\MoxicheckerKindIncrMsatQFBVStatusErrorAndUnknownCount{\the\numexpr \MoxicheckerKindIncrMsatQFBVStatusAllCount - \MoxicheckerKindIncrMsatQFBVStatusCorrectCount - \MoxicheckerKindIncrMsatQFBVStatusWrongCount \relax}
\edef\MoxicheckerKindMsatQFBVStatusErrorAndUnknownCount{\the\numexpr \MoxicheckerKindMsatQFBVStatusAllCount - \MoxicheckerKindMsatQFBVStatusCorrectCount - \MoxicheckerKindMsatQFBVStatusWrongCount \relax}
\edef\MoxicheckerKindIncrZIIIQFBVStatusErrorAndUnknownCount{\the\numexpr \MoxicheckerKindIncrZIIIQFBVStatusAllCount - \MoxicheckerKindIncrZIIIQFBVStatusCorrectCount - \MoxicheckerKindIncrZIIIQFBVStatusWrongCount \relax}
\edef\MoxicheckerKindZIIIQFBVStatusErrorAndUnknownCount{\the\numexpr \MoxicheckerKindZIIIQFBVStatusAllCount - \MoxicheckerKindZIIIQFBVStatusCorrectCount - \MoxicheckerKindZIIIQFBVStatusWrongCount \relax}
\edef\MoxiMcFlowAvrKindQFBVStatusErrorAndUnknownCount{\the\numexpr \MoxiMcFlowAvrKindQFBVStatusAllCount - \MoxiMcFlowAvrKindQFBVStatusCorrectCount - \MoxiMcFlowAvrKindQFBVStatusWrongCount \relax}
\edef\MoxiMcFlowPonoKindQFBVStatusErrorAndUnknownCount{\the\numexpr \MoxiMcFlowPonoKindQFBVStatusAllCount - \MoxiMcFlowPonoKindQFBVStatusCorrectCount - \MoxiMcFlowPonoKindQFBVStatusWrongCount \relax}
\edef\MoXIcheckerFasterCount{117}
\edef\MoXIcheckerSlowerCount{44}

\section{Introduction}
\label{sect:introduction}

\textit{Symbolic model checking}~\cite{SymbolicModelChecking,SymbolicModelChecking:McMillan1993}
embraces a wide range of automatic techniques
to formally verify a model against a specification by encoding and searching the state space symbolically.
It has been applied to hardware, software, and cyber-physical systems
to ensure their safety and correct functionality.
However, symbolic model checking has not been adopted
as widely as other ``push-button'' techniques for quality assurance, such as testing, especially in industry.
A major challenge is the lack of standardized exchange formats and
open-source implementations~\cite{MoXI-Language,CooperativeVerification}.
Even though model checkers from the same research community work on the same type of computational models,
they often use different input formats, which hinders the propagation of techniques.
Moreover, some model checkers are closed-source and
make the comparison of verification algorithms complicated,
because techniques may need to be re-implemented in a different framework
to achieve fair comparison
(this makes expensive transferability studies necessary~\cite{IMC-JAR,DAR-transferability}).

Recently, a new intermediate verification language \moxi~\cite{MoXI-Language},
the \textit{model exchange interlingua},
has been proposed to address the aforementioned challenge.
\moxi aims to be
(1)~as expressive as necessary to accommodate
real-world applications described in user-facing, higher-level modeling languages
and
(2)~as simple as possible to facilitate its translation
to tool-oriented, lower-level intermediate languages,
for which efficient and effective model checkers are available.
It augments the SMT-LIB~2~\cite{SMTLIB2} format with constructs to define state-transition systems
by using formulas in first-order logic to encode their initial and transition conditions.
\moxi inherits the expressiveness of SMT-LIB~2 and offers abundant background theories
to represent various computational models,
ranging from hardware circuits and software programs to cyber-physical systems.
The precise semantics of SMT-LIB~2 also enables
the translation from \moxi to lower-level intermediate languages.
Using SMT formulas to precisely describe state-transition systems
has also been studied in the VMT~\cite{VMT} intermediate language.

Compared to other intermediate verification languages,
such as the SMV~\cite{SMV:McMillan1993} language for finite-state transition systems
or the \boogie~\cite{Boogie} language for software programs,
using \moxi to represent model-checking problems separates
the process of encoding the semantics of computational models into SMT formulas from
the implementation of SMT-based verification approaches.
This separation will help decompose monolithic model checkers into several modular and reusable components,
e.g., standalone translators from higher-level languages to \moxi and
model-checking engines for \moxi verification tasks.
\ifdefined\techrep
A deeper discussion can be found in a recent survey on transformation for verification~\cite{TransformationGame}.
\fi

\subsection{Existing Tool Suite for \moxi}
The tool suite of \moxi~\cite{MoXI-Tool-Suite} offers translators
from SMV to \moxi and
from \moxi to the word-level modeling language \btortwo~\cite{Boolector3},
the prevailing format for hardware model checking.
The tool suite also implements a translation-based model checker, \moximcflow,
by translating a \moxi task to an equisatisfiable \btortwo task
and invoking \btortwo model checkers,
such as \avr~\cite{AVR}, \btormc~\cite{Boolector3}, and \pono~\cite{Pono},
on the translated task.
Translation-based verification approaches have been actively studied in the literature.
For example, sequential circuits in \verilog~\cite{Verilog} can be translated to SMV models~\cite{Ver2Smv,V2SMV}
or C~programs~\cite{v2c} for verification.
\btortwo circuits have been translated to C~programs and
analyzed by software verifiers~\cite{BTOR2C,Btor2-Cert,Btor2MLIR}.
C programs can also be translated to SMV or \btortwo models and
verified by hardware model checkers~\cite{CPV-SVCOMP24,Kratos2}.

While \moximcflow can solve \moxi verification tasks
by translating them to \btortwo~\cite{MoXI-Tool-Suite},
the translation-based approach limits the expressiveness of the model-checking flow
because verification problems cannot be precisely represented in \btortwo
if they involve more complex background theories,
such as integer or real arithmetics.
Moreover, to extend \moximcflow with new algorithms,
tool developers need to dig into the \btortwo model checkers.

\subsection{Motivation to Develop \moxichecker}
\label{sect:contributions}

To address the extensibility gap of the translation-based model-checking flow for \moxi,
we implemented \moxichecker, the first model checker
that solves \moxi verification tasks directly without translating them to other intermediate languages.
\moxichecker takes as input a \moxi verification task,
constructs the SMT formulas used to define the task, and
implements its verification algorithms using the API of \pysmt~\cite{PySMT},
a solver-agnostic Python library for SMT solvers.
Currently, \moxichecker supports the quantifier-free theories of bit-vectors, arrays, integers, and reals,
and the implemented algorithms include
BMC~\cite{BMC}, \kinduction~\cite{InductionVerification}, and IC3/PDR~\cite{IC3}.

The benefits of \moxichecker compared to \moximcflow are threefold.
First, \moxichecker enjoys the complete expressiveness of SMT-LIB~2
and is applicable to verification tasks involving more complex background theories,
as long as there exists an SMT solver supporting the used theory.
In contrast, \moximcflow is inadequate if the used theory is not representable
in lower-level intermediate languages focusing on bit-vectors of fixed lengths and arrays.
Second, \moxichecker allows for convenient extension and fast prototyping of model-checking algorithms.
To develop a new algorithm in \moxichecker,
one can simply work with the SMT formulas describing the model and manipulate them via the API of \pysmt.
In contrast, adding a new algorithm to the hardware model checkers used by \moximcflow
may involve encoding the semantics of hardware circuits.
Moreover, \moxichecker enables fair comparison of algorithms because
the number of confounding variables (e.g., from translation) is kept to a minimum.
Third, \moxichecker has a robust frontend design because
constructing SMT formulas that describe a \moxi verification task via \pysmt
is purely syntactical and less error-prone than translating the SMT formulas to \btortwo.

Furthermore, \moxichecker is meant for use in education.
It is an ideal framework for playing around with algorithms in course projects.
The tool has a clean architecture and a slim code base.

\inlineheadingbf{Contributions}
To sum up, our contributions in this paper include:
\begin{enumerate}
    \item \moxichecker, the first model checker that verifies \moxi tasks directly,
    \item implemented as an extensible framework to accommodate various background theories and
          facilitate the development of algorithms for \moxi,
    \item \moxichecker's first three algorithms, BMC, \kinduction, and IC3/PDR, and
    \item an evaluation of \moxichecker with \moximcflow on about 400 \moxi verification tasks.
\end{enumerate}

In our experiments, \moxichecker solved a similar number of bit-vector tasks as \moximcflow,
which used highly-optimized \btortwo model checkers as backend.
Moreover, \moxichecker was able to uniquely solve tasks using real arithmetics,
which \moximcflow cannot handle.
These contributions are significant and novel because
\moxichecker supports the standardization of symbolic model checking around \moxi
and provides an extensible framework for open-source implementations of verification algorithms for \moxi.
\section{Background}
\label{sect:background}

In this section,
we provide background knowledge for symbolic model checking
and the intermediate verification language \moxi.

\subsection{Symbolic Model Checking}
\label{sect:model-checking}

The problem of symbolic model checking~\cite{SymbolicModelChecking,SymbolicModelChecking:McMillan1993}
is to decide whether a model, usually represented as a \textit{state-transition system}~\cite{ModelLogic,HBMC-book},
satisfies a specification.
A state-transition system~$\mathcal{M}$ can be described by
an~\textit{initial condition}~$\init(s)$,
a~\textit{transition condition}~$\trans(s,s')$, and
an~\textit{invariance condition}~$\inv(s)$,
where $s$ and $s'$ range over possible states of~$\mathcal{M}$.
Condition~$\init(s)$ evaluates to $\top$ if state~$s$ is an initial state of~$\mathcal{M}$,
and $\trans(s,s')$ evaluates to $\top$ if state~$s$ can transit to state~$s'$ via one step in~$\mathcal{M}$
(we use $\top$ for $\true$).
A state~$\hat{s}$ is \textit{reachable} if
$\init(\hat{s})$ evaluates to $\top$ or
$\init(s_0) \land \trans(s_0,s_1) \land \ldots \land \trans(s_{k-1},\hat{s})$ is satisfiable for some~$k \geq 1$.
Condition~$\inv(s)$ is a constraint imposed on all reachable states in~$\mathcal{M}$
(a reachable state that violates $\inv$ is excluded for analysis).

A specification~$\varphi$ can be represented by a formula in linear temporal logic (LTL)~\cite{HBMC-TemporalLogic},
which is evaluated over the execution traces of a state-transition system.
In the following, we refer to the tuple $(\mathcal{M},\varphi)$ as a \textit{verification task},
which asks if state-transition system $\mathcal{M}$ satisfies specification $\varphi$.
\textit{Reachability safety} is an essential category of specifications,
inspecting the reachability of some target states marked by a \textit{reachable condition}~$\reachable(s)$.
A reachability-safety verification task is described by the tuple
$(\init,\trans,\inv,\reachable)$,
where $\init$, $\trans$, and $\inv$ define a state-transition system $\mathcal{M}$ and
$\reachable$ defines an LTL formula ``\textbf{always} $\lnot\reachable$''
as a specification $\varphi$ for $\mathcal{M}$.
A reachability-safety verification task is \textit{safe} (resp. \textit{unsafe})
if the target states are unreachable (resp. reachable).

In the research community of hardware model checking,
verification tasks of sequential circuits can be encoded by
the word-level language \btortwo~\cite{Boolector3}.

\subsection{The Intermediate Verification Language \moxi}
\label{sect:moxi}

\moxi~\cite{MoXI-Language}
extends the SMT-LIB~2~\cite{SMTLIB2} format with constructs to describe verification tasks.
Inheriting the expressiveness of SMT-LIB~2,
\moxi offers a variety of background theories,
ranging from bit-vectors and arrays (\qfbv and \qfabv) to
linear and nonlinear arithmetics over integers and reals (\qflia, \qflra, \qfnia, and \qfnra),
to represent models of hardware, software, and cyber-physical systems.
As for specifications, \moxi supports reachability-safety queries with fairness constraints.
We refer interested readers to the language design of \moxi~\cite{MoXI-Language} for more details.
In the following,
we use an example to show how a verification task is represented in \moxi.

\newsavebox{\exampleMoXI}
\begin{lrbox}{\exampleMoXI}
    \begin{minipage}[b]{.8\textwidth}
        \centering
\begin{lstlisting}[
            language=moxi,
            basicstyle=\ttfamilywithbold,
            numberstyle=\scriptsize,
            lineskip=2pt,
            aboveskip=0pt,
            belowskip=0pt,
            breaklines=true,
        ]
(set-logic QF_BV)(*@\label{moxi:set-logic}@*)
(define-system main(*@\label{moxi:def-sys-start}@*)
  :input ()(*@\label{moxi:def-sys-input}@*)
  :output ((s (_ BitVec 3)))(*@\label{moxi:def-sys-output}@*)
  :local ()(*@\label{moxi:def-sys-local}@*)
  :init (= s #b000)(*@\label{moxi:def-sys-init}@*)
  :trans (= s' (bvadd s #b010))(*@\label{moxi:def-sys-trans}@*)
  :inv true)(*@\label{moxi:def-sys-end}@*)
(check-system main(*@\label{moxi:check-sys-start}@*)
  :input ()(*@\label{moxi:check-sys-input}@*)
  :output ((s (_ BitVec 3)))(*@\label{moxi:check-sys-output}@*)
  :local ()(*@\label{moxi:check-sys-local}@*)
  :reachable (rch_1 (*@\newline@*)(= (bvurem s #b010) #b001))(*@\label{moxi:check-sys-reach}@*)
  :query (qry_rch_1 (rch_1)))(*@\label{moxi:check-sys-end}@*)
\end{lstlisting}
    \end{minipage}
\end{lrbox}

\Cref{fig:ex-moxi} shows a verification task of a three-bit counter in \moxi.
\Cref{moxi:set-logic} sets the background theory to \qfbv,
which allows for quantifier-free formulas over the theory of bit-vectors with fixed sizes.
\Crefrange{moxi:def-sys-start}{moxi:def-sys-end}
define the behavior of the three-bit counter with command~\texttt{define-system} and name the counter \texttt{main}.
Counter~\texttt{main} has an output variable \texttt{s},
which is a bit-vector of length three (attribute~\texttt{:output} in \cref{moxi:def-sys-output}).
Counter \texttt{main} has no inputs or local variables
(attributes~\texttt{:input} in \cref{moxi:def-sys-input}
and~\texttt{:local} in \cref{moxi:def-sys-local}, respectively).
\begin{wrapfigure}{r}{0.45\textwidth}
    \centering
    \scalebox{0.72}{\usebox{\exampleMoXI}}
    \caption{An example verification task in \moxi}
    \label{fig:ex-moxi}
    \vspace{-7mm}
\end{wrapfigure}%
The initial condition in \cref{moxi:def-sys-init} (attribute~\texttt{:init})
initializes output \texttt{s} of counter~\texttt{main} to \texttt{\#b000}.
The transition condition in \cref{moxi:def-sys-trans} (attribute~\texttt{:trans})
increments the value of \texttt{s} by \texttt{\#b010} in each step.
Note that a primed variable is treated as the next-state variable of its unprimed counterpart by \moxi.
That is, \texttt{s'} holds the value of \texttt{s} after one step.
The invariance condition in \cref{moxi:def-sys-end} (attribute~\texttt{:inv})
imposes \texttt{true} as a constraint on all reachable states of counter~\texttt{main}.
The specification for counter~\texttt{main} is described by command~\texttt{check-system}.
The reachability condition~\texttt{rch\_1} in \cref{moxi:check-sys-reach} (attribute~\texttt{:reachable})
states that the value of~\texttt{s} is an odd number,
i.e., the remainder of~\texttt{s} divided by \texttt{\#b010} equals \texttt{\#b001}.
\Cref{moxi:check-sys-end} poses a query~\texttt{qry\_rch\_1} (attribute~\texttt{:query})
to examine whether the LTL formula ``\textbf{always} $\lnot$\texttt{rch\_1}''
is satisfied by all execution traces of counter~\texttt{main}.

The \moxi tool suite~\cite{MoXI-Tool-Suite} provides an alternative representation
of \moxi verification tasks in JSON format to facilitate tool development and information exchange.
\Cref{fig:ex-json} shows the corresponding JSON file for the verification task in~\cref{fig:ex-moxi}.
Our tool \moxichecker takes \moxi verification tasks in JSON format as input.
For details of the JSON representation,
we refer interested readers to the \moxi JSON schema%
\footnote{\url{https://github.com/ModelChecker/moxi-mc-flow/tree/main/json-schema}}
in the tool suite.

\begin{figure}[t]
    \centering
    \scalebox{0.72}{
\begin{lstlisting}[
            language=moxijson,
            basicstyle=\ttfamilywithbold,
            numberstyle=\scriptsize,
            lineskip=2pt,
            aboveskip=0pt,
            belowskip=0pt,
        ]
[ { "command": "set-logic", "logic": "QF_BV" },
  { "command": "define-system",
    "symbol": "main",
    "input": [],
    "output": [{
        "symbol": "s",
        "sort": { "identifier": { "symbol": "BitVec", "indices": [3] }}}],
    "local": [],
    "init": {
      "identifier": { "symbol": "=", "indices": [] },
      "args": [{ "identifier": "s" }, { "identifier": "#b000" }]},
    "trans": {
      "identifier": { "symbol": "=", "indices": [] },
      "args": [
        { "identifier": "s'" },
        {"identifier": { "symbol": "bvadd", "indices": [] },
          "args": [{ "identifier": "s" }, { "identifier": "#b010" }]}]},
    "inv": { "identifier": "true" }},
  { "command": "check-system",
    "symbol": "main",
    "input": [],
    "output": [{
        "symbol": "s",
        "sort": { "identifier": { "symbol": "BitVec", "indices": [3] }}}],
    "local": [],
    "reachable": [
      { "symbol": "rch_1",
        "formula": { "identifier": { "symbol": "=", "indices": [] },
          "args": [
            {"identifier": { "symbol": "bvurem", "indices": [] },
              "args": [{ "identifier": "s" }, { "identifier": "#b010" }]},
            { "identifier": "#b001" }]}}],
    "query": [{ "symbol": "qry_rch_1", "formulas": ["rch_1"] }]}]
\end{lstlisting}
    }
    \vspace{-2mm}
    \caption{A JSON representation of the \moxi verification task in~\cref{fig:ex-moxi}}
    \label{fig:ex-json}
    \vspace{-2mm}
\end{figure}

To analyze a \moxi verification task,
the model checker \moximcflow in the \moxi tool suite
translates the \moxi task to an equisatisfiable \btortwo verification task
and invokes hardware model checkers for \btortwo, e.g.,
\avr~\cite{AVR},
\btormc~\cite{Boolector3}, and
\pono~\cite{Pono},
from the Hardware Model Checking Competitions~\cite{HWMCC17}.
\section{Software Architecture of \moxichecker}
\label{sect:architecture}

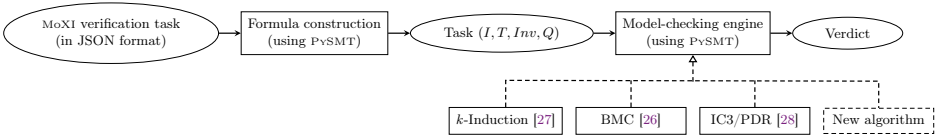
\begin{figure}[t]
    \centering
    \scalebox{.58}{\begin{tikzpicture}[every node/.style={font=\small}]
    \tikzstyle{data} = [ellipse, minimum width=2.5cm, minimum height=5mm, inner sep=1.5mm, text centered, draw=black, align=center]
    \tikzstyle{process} = [rectangle, minimum width=2.5cm, minimum height=5mm, inner sep=1.5mm, text centered, draw=black, align=center]
    \tikzstyle{arrow} = [thick, ->, >=stealth]
    \tikzstyle{implement} = [arrow, densely dashed, -{Triangle[open]}]

    \node (moxi) [data] {\moxi verification task\\(in JSON format)};
    \node (frontend) [process, right=5mm of moxi] {Formula construction\\(using \pysmt)};
    \node (formulas) [data, right=5mm of frontend] {Task~$(\init,\trans,\inv,\reachable)$};
    \node (model-checking) [process, right=5mm of formulas] {Model-checking engine\\(using \pysmt)};
    \node (bmc) [process, below left=1.2cm and 1.5mm of model-checking.south] {BMC~\cite{BMC}};
    \node (ki) [process, left=3mm of bmc] {\kInduction~\cite{InductionVerification}};
    \node (ic3) [process, right=3mm of bmc] {IC3/PDR~\cite{IC3}};
    \node (new-alg) [process, densely dashed, right=3mm of ic3] {New algorithm};
    \node (verdict) [data, right=5mm of model-checking] {Verdict};
    \node (mce-conn) [below=5mm of model-checking] {};
    \node (bmc-conn) [above=5mm of bmc] {};
    \node (ki-conn) [above=5mm of ki] {};
    \node (ic3-conn) [above=5mm of ic3] {};
    \node (newa-conn) [above=5mm of new-alg] {};

    \draw [arrow] (moxi) -- (frontend);
    \draw [arrow] (frontend) -- (formulas);
    \draw [arrow] (formulas) -- (model-checking);
    \draw [arrow] (model-checking) -- (verdict);
    \draw [implement] (mce-conn.center) -- (model-checking.south);
    \draw [thick, densely dashed] (ki.north) -- (ki-conn.center) -- (newa-conn.center) -- (new-alg.north);
    \draw [thick, densely dashed] (bmc.north) -- (bmc-conn.center);
    \draw [thick, densely dashed] (ic3.north) -- (ic3-conn.center);
\end{tikzpicture}}
    \vspace{-6mm}
    \caption{Software architecture of \moxichecker}
    \label{fig:architecture}
    \vspace{-2mm}
\end{figure}

\Cref{fig:architecture} shows the software architecture of \moxichecker,
the first model checker for \moxi without translating verification tasks to lower-level languages.
Implemented in the programming language Python,
\moxichecker is open-source on GitLab%
\ifdefined\techrep
    \footnote{\url{\moxicheckerurl}}
\else
    {}
\fi
and released under the Apache License 2.0.
On a \moxi verification task in JSON format,
\moxichecker uses the standard JSON package of Python to load the input file
and constructs SMT formulas for the initial, transition, invariance, and reachable conditions
by calling the API of the solver-agnostic library \pysmt~\cite{PySMT} for SMT solvers.
It then performs model checking on
the reachability-safety verification task $(\init,\trans,\inv,\reachable)$.
The output of \moxichecker on a \moxi verification task is a verdict
to indicate whether the task is safe or unsafe.

Different from \moximcflow in the \moxi tool suite~\cite{MoXI-Tool-Suite},
which translates verification tasks in \moxi to \btortwo~\cite{Boolector3} and invokes hardware model checkers,
\moxichecker implements its model-checking engines using the API of \pysmt.
Currently, \moxichecker supports
\texttt{QF\_BV},
\texttt{QF\_ABV},
\texttt{QF\_LIA},
\texttt{QF\_LRA},
\texttt{QF\_NIA}, and
\texttt{QF\_NRA} as the background theory.
The elegant software architecture facilitates adding new background theories to \moxichecker.

We adapted and integrated the implementations of
BMC~\cite{BMC}, \kinduction~\cite{InductionVerification}, and IC3/PDR~\cite{IC3} in \pysmt%
\footnote{\url{https://github.com/pysmt/pysmt/blob/master/examples/model_checking.py}} into our framework.
In addition, we demonstrate the extensibility of \moxichecker by contributing a \kinduction implementation
that takes advantage of incremental solving of SMT solvers by reusing solver stacks.
Compared to the non-incremental version in \pysmt,
the incremental \kinduction was more efficient and solved more tasks in the evaluation.

\subsection{Example}
\label{sect:example}

We demonstrate the working of \moxichecker by invoking it on the verification task in~\cref{fig:ex-moxi}.
\moxichecker consumes the JSON file of the \moxi verification task in~\cref{fig:ex-json} as input and
constructs SMT formulas $s=0$, $s'=s+2$, $\top$, and $s \% 2 = 1$,
as the initial, transition, invariance, and reachability conditions, respectively.
Note that variable~$s$ is a bit-vector of length three, and variable~$s'$ is its next-state counterpart.
To honor the invariance condition,
\moxichecker conjoins it with initial and transition conditions, respectively.
As the invariance condition is $\top$ in the example verification task,
we omit it in the following explanation.

To solve the verification task,
\moxichecker considers ``\textbf{always} $\lnot (s \% 2 = 1)$''
as the specification for the state-transition system.
By applying \kinduction~\cite{InductionVerification},
\moxichecker shows that both
the \textit{base case} $(s=0) \implies \lnot (s \% 2 = 1)$ and
the \textit{step case} $\lnot (s \% 2 = 1) \land (s'=s+1) \implies \lnot (s' \% 2 = 1)$ for $k=1$ hold.
Therefore, \moxichecker concludes that counter~\texttt{main} in~\cref{fig:ex-moxi} satisfies its specification.

\subsection{Current Limitations}
\label{sect:limitations}

\moxi is an expressive and versatile intermediate language for describing verification tasks.
The current version of \moxichecker
\ifdefined\techrep
    (release~\href{\moxicheckerurl/-/tree/0.1}{0.1})
\fi
still misses some language support for \moxi.
For instance, \moxi allows the composition of multiple state-transition systems
via attribute \texttt{:subsys} in command \texttt{define-system}
and fairness constraints via attribute \texttt{:fairness} in command \texttt{check-system}.
In addition, \moxi defines a format for \emph{verification witnesses}~\cite{WitnessesJournal,CertifyingAlgorithms},
e.g., an error trace if the specification is violated
or an invariant if the specification is satisfied.
We are actively working on supporting these language features
to make \moxichecker more comprehensive.
\pgfplotsset{quantile plot/.style={
    width=6cm,
    height=3.5cm,
    scale only axis,
    /pgfplots/table/x expr={\coordindex+1},
    /pgfplots/table/y index=4,
    /pgfplots/table/header=false,
    ylabel shift=-1em,
    ticklabel style={font={\smaller}},
    xmin=0,
    ymin=1,
    ymax=1000,
    legend cell align={left},
    legend style={at={(0,1)}, anchor=north west, outer xsep=5pt, outer ysep=5pt, fill=none, font={\smaller}},
    legend columns=3,
    cycle multiindex list={
      orange, green, blue, purple, teal, brown, black\nextlist
      mark list*\nextlist
      solid, densely dashed, densely dashdotdotted, densely dotted},
  },
  every axis plot/.append style={thick}
}

\onlyifstandalone{
  \pgfplotsset{quantile plot/.append style={
      ylabel=CPU time (\second),
    },
  }
}

\newcommand\addgraph[2]{{
  \newcommand\csvfile{\plotpath/\detokenize{#2}}
  \IfFileExists\csvfile{
    \addplot+ table {\csvfile}; \addlegendentry{#1}
  }{
    \addplot coordinates {};
  }
}}

\newcommand{\numbvtasks}{\num{\MoxicheckerKindIncrMsatQFBVStatusAllCount}\xspace}
\newcommand{\numbvtruetasks}{\num{225}\xspace}
\newcommand{\numbvfalsetasks}{\num{157}\xspace}
\newcommand{\numothertasks}{\num{9}\xspace}
\newcommand{\numuniqsolvedtasks}{\num{17}}
\newcommand{\numbothsolvedtasks}{\num{161}}

\section{Evaluation}
\label{sect:evaluation}

To demonstrate the performance and extensibility of \moxichecker,
we compared it to \moximcflow,
the translation-based model checker for \moxi~\cite{MoXI-Tool-Suite},
which invokes hardware model checkers for \btortwo as backend.
Our experiments aim to answer the following research questions:

\begin{itemize}
    \item RQ1: Is \moxichecker effective and efficient compared to \moximcflow on \qfbv tasks?
    \item RQ2: Can \moxichecker solve tasks using more complex background theories,
          which \moximcflow cannot solve?
\end{itemize}

\subsection{Experimental Setup}
\label{sect:exp-setup}

We evaluated \moxichecker and \moximcflow on two sets of \moxi verification tasks in JSON format.
The first benchmark set consists of \numbvtruetasks~safe and \numbvfalsetasks~unsafe \qfbv tasks,
taken from the \moxi tool suite~\cite{MoXI-Tool-Suite}.
Due to the lack of publicly available verification tasks involving more complex theories,
we handcrafted \numothertasks~tasks using the theories of \qflia, \qflra, \qfnia, and \qfnra
to show the extensible theory support of \moxichecker.

We used \moxichecker
\ifdefined\techrep
\href{\moxicheckerurl/-/tree/0.1}{version~0.1}
\else
version~0.1
\fi
and \moximcflow at commit
\href{https://github.com/ModelChecker/moxi-mc-flow/tree/6240207d1ae5629751b15ccddba97b2021e8d7f7}{6240207d}
in the experiments.
\moxichecker called SMT solvers \zthree~\cite{Z3} and \mathsat~\cite{MATHSAT5}
for \qfbv, \qflia, and \qflra tasks;
for tasks using nonlinear arithmetics, \moxichecker employed \zthree.
\moximcflow invoked \btortwo model checkers \avr~\cite{AVR} and \pono~\cite{Pono} to solve \qfbv tasks;
\qflia and \qfnia tasks were also solved by \btortwo model checkers via encoding integers with 32 bits.
The version of \moximcflow used in our evaluation had no support for reals.
Both \moxichecker and \moximcflow used \kinduction for verification.
(For \moximcflow, \avr and \pono were configured to use \kinduction on translated \btortwo tasks.)

All experiments were conducted on a machine that runs a
GNU/Linux operating system (x86_64-linux, Ubuntu 22.04 with Linux kernel 5.15)
and is equipped with \SI{2}{TB} of RAM in total and
two \SI{2.0}{GHz} AMD EPYC 7713 CPUs with 128~processing units each.
Each task was limited to 2~CPU cores,
\SI{15}{min} of CPU time,
and \SI{15}{GB} of RAM.
We used \benchexectoolurl~\cite{Benchmarking-STTT}
to ensure reliable resource measurement and reproducible results.

\subsection{Experimental Results}
\label{sect:exp-results}

\inlineheadingbf{RQ1: Performance of \moxichecker}
\Cref{tab:bv-summary} summarizes the experimental results of \moxichecker and \moximcflow
on \numbvtasks~\qfbv verification tasks.
\moxichecker, when using \mathsat as the backend solver and incremental solving,
delivered the most correct results.
Notably, \moxichecker solved \numuniqsolvedtasks~tasks that \moximcflow failed to translate to \btortwo.

Despite being implemented in Python,
\moxichecker demonstrated a comparable performance to \moximcflow,
which employs highly-optimized hardware model checkers written in \CC as backend.
This is mainly because the bottleneck of SMT-based verification algorithms lies in solving SMT formulas.
A preliminary run-time profiling for \moxichecker by
\href{https://docs.python.org/3/library/profile.html}{\tool{cProfile}}
showed that solving formulas accounted for more than \SI{90}{\%} of the run-time
for the more time-consuming tasks.
The results suggest that using Python to construct and manipulate SMT formulas
does not incur much overhead for \moxichecker.

\begin{table}[t]
    \centering
    \caption{Summary of verification results on \numbvtasks \qfbv tasks}
    \label{tab:bv-summary}
    \begin{tabular}{l@{\hspace{2mm}}|@{\hspace{2mm}}S[table-format=3]@{\hspace{2mm}}S[table-format=3]@{\hspace{2mm}}S[table-format=3]@{\hspace{2mm}}S[table-format=3]@{\hspace{2mm}}|@{\hspace{2mm}}S[table-format=3]@{\hspace{2mm}}S[table-format=3]}
        \toprule
        Tool                & \multicolumn{4}{c|@{\hspace{2mm}}}{\moxichecker}                & \multicolumn{2}{c}{\moximcflow}                                                              \\
        Backend             & \msat                                                           & \msat{}$\textsuperscript{incr}$ & \zthree & \zthree{}$\textsuperscript{incr}$ & \avr & \pono \\
        \midrule
        Correct results     & \MoxicheckerKindMsatQFBVStatusCorrectCount
                            & {\bfseries \MoxicheckerKindIncrMsatQFBVStatusCorrectCount}
                            & \MoxicheckerKindZIIIQFBVStatusCorrectCount
                            & \MoxicheckerKindIncrZIIIQFBVStatusCorrectCount
                            & \MoxiMcFlowAvrKindQFBVStatusCorrectCount
                            & \MoxiMcFlowPonoKindQFBVStatusCorrectCount
        \\
        \quad Proofs        & \MoxicheckerKindMsatQFBVStatusCorrectTrueCount
                            & \MoxicheckerKindIncrMsatQFBVStatusCorrectTrueCount
                            & {\bfseries \MoxicheckerKindZIIIQFBVStatusCorrectTrueCount}
                            & {\bfseries \MoxicheckerKindIncrZIIIQFBVStatusCorrectTrueCount}
                            & \MoxiMcFlowAvrKindQFBVStatusCorrectTrueCount
                            & \MoxiMcFlowPonoKindQFBVStatusCorrectTrueCount
        \\
        \quad Alarms        & \MoxicheckerKindMsatQFBVStatusCorrectFalseCount
                            & {\bfseries \MoxicheckerKindIncrMsatQFBVStatusCorrectFalseCount}
                            & \MoxicheckerKindZIIIQFBVStatusCorrectFalseCount
                            & \MoxicheckerKindIncrZIIIQFBVStatusCorrectFalseCount
                            & \MoxiMcFlowAvrKindQFBVStatusCorrectFalseCount
                            & \MoxiMcFlowPonoKindQFBVStatusCorrectFalseCount
        \\
        Errors and Unknown & \MoxicheckerKindMsatQFBVStatusErrorAndUnknownCount
                            & \MoxicheckerKindIncrMsatQFBVStatusErrorAndUnknownCount
                            & \MoxicheckerKindZIIIQFBVStatusErrorAndUnknownCount
                            & \MoxicheckerKindIncrZIIIQFBVStatusErrorAndUnknownCount
                            & \MoxiMcFlowAvrKindQFBVStatusErrorAndUnknownCount
                            & \MoxiMcFlowPonoKindQFBVStatusErrorAndUnknownCount
        \\
        \bottomrule
    \end{tabular}
\end{table}

\begin{figure}[t]
    \centering
    \scalebox{.95}{\begin{tikzpicture}
\begin{semilogyaxis}[
    xlabel=n-th fastest correct result,
    ylabel=CPU time (\second),
    quantile plot,
    width=10cm,
    height=4cm,
    mark repeat=25,
    /pgf/number format/1000 sep={\,},
    legend style={font=\tiny, at={(0, 1)}, anchor=north west},
    legend columns=1,
    xmax=180,
    xmin=0,
    ymin=0
    ]
    \addgraph{\moxichecker{}$\cdot$\msat}{../csv/moxichecker.kind-msat.QF_BV.quantile.csv}
    \addgraph{\moxichecker{}$\cdot$\msat{}$\textsuperscript{incr}$}{../csv/moxichecker.kind-incr-msat.QF_BV.quantile.csv}
    \addgraph{\moximcflow{}$\cdot$\avr}{../csv/moxi-mc-flow.avr-kind.QF_BV.quantile.csv}
    \addgraph{\moximcflow{}$\cdot$\pono}{../csv/moxi-mc-flow.pono-kind.QF_BV.quantile.csv}
\end{semilogyaxis}
\end{tikzpicture}}
    \vspace{-3mm}
    \caption{\moxichecker vs. \moximcflow on \numbvtasks \qfbv tasks}
    \label{fig:quantile}
    \vspace{-2mm}
\end{figure}

\begin{figure}
    \centering
    \begin{minipage}{.48\textwidth}
        \centering
        \scalebox{.7}{\begin{tikzpicture}
\begin{loglogaxis}[
    xlabel=\moximcflow{}$\cdot$\avr (\second),
    ylabel=\moxichecker{}$\cdot$\msat{}$\textsuperscript{incr}$ (\second),
    xmin=0.1,
    xmax=1000,
    ymin=0.1,
    ymax=1000,
    domain=0.1:1001,
    clip mode=individual,
    axis equal image,
    legend pos=north west,
    /pgf/number format/1000 sep={\,},
    ]
    \addplot+[green, mark=o, only marks, opacity=0.3]
        table[
            header=false,
            skip first n=3, 
            x index=5, 
            y index=3, 
            ] {eval-results/csv/moxichecker-mcflow.cputime.scatter.table.csv};
    \addplot[gray] {x};
    \addplot[gray] {10*x};
    \addplot[gray] {x/10};
\end{loglogaxis}
\end{tikzpicture}}
        \vspace{-3mm}
        \caption{Efficiency of \moxichecker vs. \moximcflow on \qfbv tasks}
        \label{fig:scatter-cputime}
    \end{minipage}
    \hfill
    \begin{minipage}{.48\textwidth}
        \centering
        \scalebox{.7}{\begin{tikzpicture}
\begin{loglogaxis}[
    xlabel=\moxichecker{}$\cdot$\msat{} (\second),
    ylabel=\moxichecker{}$\cdot$\msat{}$\textsuperscript{incr}$ (\second),
    xmin=0.1,
    xmax=1000,
    ymin=0.1,
    ymax=1000,
    domain=0.1:1001,
    clip mode=individual,
    axis equal image,
    legend pos=north west,
    /pgf/number format/1000 sep={\,},
    ]
    \addplot+[green, mark=o, only marks, opacity=0.3]
        table[
            header=false,
            skip first n=3, 
            x index=5, 
            y index=3, 
            ] {eval-results/csv/incr-solving.cputime.scatter.table.csv};
    \addplot[gray] {x};
    \addplot[gray] {10*x};
    \addplot[gray] {x/10};
\end{loglogaxis}
\end{tikzpicture}}
        \vspace{-3mm}
        \caption{Effect of incremental SMT solving in \moxichecker on \qfbv tasks}
        \label{fig:scatter-incr}
    \end{minipage}
    \vspace{-2mm}
\end{figure}

In our evaluation, \moxichecker was also more efficient than \moximcflow in terms of CPU-time consumption.
\Cref{fig:quantile} shows a quantile plot
comparing \moxichecker and \moximcflow on the \qfbv tasks.
A data point $(x, y)$ in the plot indicates that there are $x$ tasks,
each of which can be correctly solved by the respective tool within a time bound $y$ seconds.
The figure shows that \moxichecker ran faster than \moximcflow,
especially for tasks that can be solved quickly,
because \moximcflow had a slower startup time due to its translation process
(note the higher y-intercept of roughly \SI{3}{s} in \cref{fig:quantile}).

\Cref{fig:scatter-cputime} shows a head-to-head comparison of
\moxichecker (cf.\,~\textcolor{green}{\raisebox{2.5pt}{\pgfuseplotmark{square*}}}~~in \cref{fig:quantile})
and
\moximcflow (cf.\,~\textcolor{blue}{\raisebox{2.5pt}{\pgfuseplotmark{triangle*}}}~~in \cref{fig:quantile})
in a scatter plot.
A data point $(x, y)$ in the plot represents a task that was solved by both \moximcflow and \moxichecker,
for which the former took x seconds, while the latter took y seconds.
The figure shows that the efficiency of \moxichecker was competitive against \moximcflow.
In particular, out of the \numbothsolvedtasks~tasks solved by both,
\moxichecker was faster than \moximcflow on \MoXIcheckerFasterCount~tasks.

In addition to the comparison with \moximcflow,
we evaluated the impact of backend solvers and incremental solving on \moxichecker.
From \cref{tab:bv-summary}, observe that \mathsat and \zthree delivered similar performance,
with the former being slightly more effective.
In contrast, incremental SMT solving had a more pronounced effect
on both the effectiveness and efficiency of \moxichecker.
The performance improvement of our \kinduction implementation (\moxichecker{}$\cdot$\msat{}$\textsuperscript{incr}$)
over the implementation provided by \pysmt (\moxichecker{}$\cdot$\msat{})
is also evident in \cref{fig:quantile} and \cref{fig:scatter-incr}.

\inlineheadingbf{RQ2: Extensibility of \moxichecker}
\Cref{tab:theories} lists the results of \moxichecker and \moximcflow
on \numothertasks~handcrafted model-checking problems involving integer and real arithmetics.
\moxichecker correctly solved all tasks.
In contrast, \moximcflow produced
wrong results or timeouts for the tasks containing integers (upper half of \cref{tab:theories})
and had no support for tasks containing reals (lower half of \cref{tab:theories}).
Unlike \moxichecker, which utilized \zthree and thus supported the theories over integers and reals,
\moximcflow approximated integers with bit-vectors (of length 32 by default).
Due to the potential issues of overflow and underflow in bit-vector arithmetics,
such approximation is both \textit{unsound} and \textit{incomplete},
therefore causing the incorrect verification results in \cref{tab:theories}.
This illustrative experiment shows that, compared to \moximcflow,
\moxichecker is
(1) more reliable, as it does not yield wrong results due to approximation, and
(2) more versatile, as it supports many background theories.

\begin{table}[t]
    \centering
    \caption{\moxichecker vs. \moximcflow on tasks using integers and reals}
    \label{tab:theories}
    \begin{tabular}{c@{\hspace{2.5mm}}c@{\hspace{2.5mm}}c@{\hspace{2.5mm}}|@{\hspace{2.5mm}}c@{\hspace{2.5mm}}c}
        \toprule
        Task                                                              & Theory & Verdict & \moxichecker & \moximcflow             \\
        \midrule
        \moxiextask{QF_LIA/FibonacciSequence_unreach2}{FibonacciSequence} & \qflia & safe    & safe         & \textcolor{red}{unsafe} \\
        \moxiextask{QF_LIA/IntIncrement_reach}{IntIncrement}              & \qflia & unsafe  & unsafe       & \textcolor{red}{safe}   \\
        \moxiextask{QF_LIA/IntCounter_unreach}{IntCounter}                & \qflia & safe    & safe         & timeout                 \\
        \moxiextask{QF_NIA/IntMultiply}{IntMultiply}                      & \qfnia & safe    & safe         & \textcolor{red}{unsafe} \\
        \midrule
        \moxiextask{QF_LRA/BoundedLinearGrowth}{BoundedLinearGrowth}      & \qflra & safe    & safe         & unsupported             \\
        \moxiextask{QF_LRA/DoubleDelay2}{DoubleDelay2}                    & \qflra & unsafe  & unsafe       & unsupported             \\
        \moxiextask{QF_NRA/OscillatingRatio}{OscillatingRatio}            & \qfnra & safe    & safe         & unsupported             \\
        \moxiextask{QF_NRA/SafeNonlinearGrowth}{SafeNonlinearGrowth}      & \qfnra & safe    & safe         & unsupported             \\
        \moxiextask{QF_NRA/NonlinearGrowth}{NonlinearGrowth}              & \qfnra & unsafe  & unsafe       & unsupported             \\
        \bottomrule
    \end{tabular}
\end{table}

\section{Conclusion}
\label{sect:conclusion}

We introduced \moxichecker, the first model checker for \moxi
that performs model checking with the SMT formulas describing a \moxi task directly.
Compared to \moximcflow~\cite{MoXI-Tool-Suite},
which translates verification tasks to \btortwo~\cite{Boolector3} and invokes hardware model checkers,
\moxichecker accommodates \moxi verification tasks with various background theories,
facilitates the implementation of new model-checking algorithms,
abstracts from specific SMT solvers using the API of \pysmt,
and has a robust frontend design that avoids potential translation bugs.
Currently, \moxichecker supports the quantifier-free theories of bit-vectors, arrays, integers, and reals,
and implements BMC~\cite{BMC}, \kinduction~\cite{InductionVerification}, and IC3/PDR~\cite{IC3} for verification.
In our evaluation, \moxichecker achieved a comparable performance against \moximcflow on bit-vector tasks
and uniquely solved tasks using integer or real arithmetics.
We envision \moxichecker to facilitate open-source implementations for model-checking techniques around \moxi
and become a cornerstone for wider adoption of symbolic model checking.
For future work, we will enhance the language support of \moxichecker,
improve the existing verification algorithms and implement new ones,
and apply \moxichecker to software programs or cyber-physical systems.

\subsubsection{Data-Availability Statement}
\ifdefined\techrep
The \moxichecker release 0.1
is available at Zenodo~\cite{MOXIchecker-artifact-VSTTE24-submission} and at \url{\moxicheckerurl}.
\else
\moxichecker's source code and executables 
are available at Zenodo~\cite{MOXIchecker-artifact-VSTTE24-submission}.
\fi

\ifdefined\techrep
\subsubsection{Funding Statement}
This project was funded in part by the Deutsche Forschungsgemeinschaft (DFG)
-- \href{http://gepris.dfg.de/gepris/projekt/378803395}{378803395} (ConVeY)
and \href{http://gepris.dfg.de/gepris/projekt/536040111}{536040111} (Bridge).
\fi

\def\UrlBigBreaks{\do\/\do-\do:}
\interlinepenalty=10000
\bibliography{bibs/dbeyer,bibs/artifacts,bibs/sw,bibs/svcomp,bibs/svcomp-artifacts,bibs/websites}





\end{document}